\documentclass[preprint,showpacs,preprintnumbers,amsmath,amssymb]{revtex4}

\usepackage{graphicx}
\usepackage{dcolumn}
\usepackage{bm}
\usepackage{amsfonts}%
\usepackage{amsmath}

\begin{document}


\title{Photofragmentation of the H$_3$ molecule, including Jahn-Teller coupling effects}

\author{Viatcheslav Kokoouline\footnote{Present address: Department of Physics, University of Central Florida, Orlando, FL 32816-2385, USA} and Chris H. Greene}
\affiliation{Department of Physics and JILA, University of Colorado, Boulder, Colorado 80309-0440, USA}
\date{\today}

\begin{abstract}
We have developed a theoretical  method for interpretation of photoionization experiments with  the H$_3$ molecule. In the present study we give a detailed description of the method, which combines multichannel quantum defect theory, the adiabatic hyperspherical approach, and the techniques of outgoing Siegert pseudostates. The present method accounts for vibrational and rotation excitations of the molecule, deals with  all symmetry restrictions imposed by the geometry of the molecule, including vibrational, rotational, electronic and nuclear spin symmetries. The method was recently applied to treat dissociative recombination of the H$_3^+$ ion. Since  H$_3^+$ dissociative recombination has been a controversial problem, the present study also allows us to test the method on the process of photoionization, which is understood better than dissociative recombination.  Good agreement with two photoionization experiments is obtained.
\end{abstract}

\vspace*{0cm}

\maketitle

\section{Introduction}

The simplest polyatomic molecules, H$_3$ and H$_3^+$, have been intensively studied during the last decades. Interest in these molecules is motivated by the fact that the H$_3^+$ ion plays an important role in the chain of chemical reactions in interstellar space, acting as protonator in chemical reactions with almost all atoms. In particular, dissociation recombination (DR) of H$_3^+$ with an electron leads via several intermediate steps to the production of water in interstellar space. Many successful models in interstellar chemistry are based on  H$_3^+$ DR. In addition, H$_3^+$ attracts theorists as a benchmark for high accuracy calculations with small molecules: Theoretical {\it ab initio } methods can be tested against existing experimental H$_3^+$  spectroscopy data. The interest in the neutral H$_3$ molecule  is closely related to the problem of H$_3^+$ DR. But H$_3$ also presents great interest from another point of view. Experimental studies \cite{helm86,helm88,dodhy88,bordas91,mistrik00} of the metastable  H$_3$ molecule and  several later theoretical studies \cite{bordas91,mistrik00,stephens94,stephens95} revealed  non-Born-Oppenheimer effects of coupling between its electronic, vibrational, and rotational degrees of freedom. As was shown recently \cite{kokoouline03a,kokoouline03b}, these non-Born-Oppenheimer effects play an important role in H$_3^+$ DR as well. The present study is devoted to a theoretical treatment of H$_3$ photoionization. 

There are three main reasons for this study. The first one is to explore a new theoretical method for the treatment of polyatomic photoionization. Our method is based on multi-channel quantum defect theory (MQDT) \cite{seaton83,fano86,aymar96,jungen96}, the adiabatic hyperspherical approach to vibrational dynamics of three nuclei, the formalism of outgoing wave Siegert states \cite{tolstikhin97,tolstikhin98,hamilton02}, and inclusion of a non-Born-Oppenheimer coupling---Jahn-Teller effect. The second reason is related to a recent study of H$_3^+$ DR \cite{kokoouline03a,kokoouline03b}, where the reported method was very successful in treatment of H$_3^+$ DR, giving good agreement between theoretical calculations and experimental results from storage rings \cite{mccall03,tanabe00,jensen01}. However, since that method is new, it is desirable to test it in greater detail. An application to the interpretation of H$_3$ photoionization experiments \cite{bordas91,mistrik00} is such a test. These photoionization experiments were successfully interpreted in previous theoretical work \cite{bordas91,mistrik00,stephens94,stephens95}, where another method  based on MQDT was applied. Thus, the present treatment can also be tested against the previous theoretical studies.

Our treatment of photoionization is similar to the one developed by Stephens and Greene \cite{stephens94,stephens95}, and employed in Refs. \cite{mistrik00,stephens94,stephens95} for interpretation of two photoionization experiments by Bordas {\it et al.} \cite{bordas91} and by Mistr\'\i k {\it et al.} \cite{mistrik00}. Both experiments were  interpreted  using a full rovibronic  frame transformation \cite{stephens94,stephens95}. The present treatment has several differences from the one proposed by  Stephens and Greene. The first difference  is the use of the adiabatic hyperspherical approximation  \cite{zhou93,lin95,esry96} for the representation of vibrational wave functions.   Stephens and Greene used the exact three-dimensional vibrational wave functions. The second difference is the correction of the incompatibility  between the form for the reaction matrices used in Refs. \cite{stephens94,stephens95,mistrik00,kokoouline01} and the quantum defect parameters of Jahn-Teller coupling used in the mentioned studies. In fact, the  values of  Jahn-Teller quantum defect parameters  used in Refs. \cite{mistrik00,stephens94,stephens95,kokoouline01} are compatible with an alternative form of the reaction matrix, which was adopted in Refs. \cite{longuet61,staib90a,staib90b}. In the present work we use the same form of $K$-matrix as in Refs. \cite{mistrik00,stephens94,stephens95,kokoouline01} and quantum defect parameters from Ref. \cite{mistrik00}. Thus, Jahn-Teller parameters $\delta$ and $\lambda$ from  \cite{mistrik00} should be multiplied by $-\pi$ to be used in the present study.  The third difference is in the symmetrization of the total rovibrational wave functions of the H$_3^+$ ion. In Refs.  \cite{mistrik00,stephens94,stephens95} the symmetrization is made according to the procedure proposed by Spirko and Jensen \cite{spirko85}:  Rotational and vibrational parts of the total wave function are symmetrized  separately and in two-step procedure. In the present treatment we symmetrize the total wave function only once at the very final step. This greatly simplifies the construction of wave functions of a required symmetry. The fourth difference is in calculation of dipole transition moments. Calculating the dipole moment into a final state, Stephens and Greene accounted only for the diagonal component of the final state wave function. Our treatment accounts for all non-diagonal wave function components contributing to the dipole transition element.

The article is organized as follows. 
Section \ref{sec:tot_wf} describes construction of the total wave function of H$_3^+$ and compares our method of the construction with the method proposed in Ref. \cite{spirko85}. In Sec. \ref{sec:S}, we build up the scattering matrix that represents the collision between an electron and an ion.  Section \ref{sec:oscil_strength} presents a derivation of dipole transition moments and oscillator strengths for H$_3$. We discuss results of our calculation and compare those results with experimental data in Sec. \ref{sec:interpretation_of_spectr}. Section \ref{sec:concl} states our conclusions. 

Atomic units are used in the article unless otherwise stated.

\section{Symmetry of the total wave function of H$_3^+$}
\label{sec:tot_wf}

In this study we consider only $p$-wave scattering  (or half-scattering) of the electron from the molecule. As demonstrated in Refs. \cite{mistrik00,stephens95}, higher electronic partial waves make much smaller contribution  than the $p$-wave to the photoionization spectrum. Similar to our study of H$_3^+$ dissociative recombination, we chose the  molecular axis $Z$ along the main symmetry axis of the molecule. Directions of two other axes, $X$ and $Y$, are shown in Fig. 3 of Ref. \cite{kokoouline03b}.

\subsection{Total wave function}

The total wave function $\Phi_{t}^{n.sym}$ of the ion can be represented as a sum of terms, each of which is product of three factors \cite{kokoouline03b}:
\begin{equation}
\label{eq:phi_total}
\Phi_{t}^{n.sym}=\Phi^I_{g_I} {\cal R}_{N^+K^+m^+}(\alpha , \beta , \gamma) \Phi_{\mathbf v}({\cal Q}).
\end{equation}
In the above equation, $\alpha , \beta$, and $\gamma$ are three Euler angles defining the orientation of the molecular fixed axis with respect to the space fixed coordinates system.  Below, we describe briefly the construction of all three factors in the product of Eq. (\ref{eq:phi_total}). A more detailed description is given in Ref. \cite{kokoouline03b}. 

The rotational part ${\cal R}(\alpha , \beta , \gamma)$ of the total wave function in Eq. (\ref{eq:phi_total}) is the symmetric top wave function for H$_3^+$, which is proportional to the Wigner function \cite{bunkerbook}. The quantum numbers $N^+$, $K^+$, and $m^+$ refer to the total angular momentum $N^+$ and its projections on the molecular $Z$-axis, $K^+$, and the laboratory $z$-axis, $m^+$.  The transformation properties  of the symmetric top wave function under the  $D_{3h}$ group, are  given in Table II of Ref. \cite{kokoouline03b}.

The vibrational symmetry of H$_3^+$ and H$_3$ is described by the group $C_{3v}$.  $C_{3v}$ is a subgroup of $D_{3h}$: $D_{3h}=\sigma_h\otimes C_{3v}$, where $\sigma_h$ is the operation of reflection with respect to the plane of three nuclei.  For our discussion of the vibrational symmetry of H$_3^+$, it is convenient to use normal coordinates $Q_1$, $Q_x$, and $Q_y$ (for definitions, see, for example, Ref. \cite{mistrik00}). $Q_1$  describes the symmetric  stretch mode. The motion along this coordinate is characterized by the (approximate) quantum number $v_1$ and by the corresponding frequency $\omega_1$. Normal coordinates $Q_x$ and $Q_y$  correspond to two vibrational modes having the same frequency of oscillations $\omega_2$. Vibrations along  $Q_x$ and $Q_y$  are characterized by the approximate numbers $v_x$ and $v_y$, correspondingly. The total vibrational energy can be approximated $E=\omega_1(v_1+1/2)+\omega_2(v_x+v_y+1)$. (The vibrational quantum numbers $v_1$, $v_x$ and $v_y$ and the corresponding energy $E$ are not exact as long as the ionic molecular potential is not exactly harmonic.)  Due to the degeneracy of the $Q_x$ and $Q_y$ modes, the two-dimensional vibrational motion along the $Q_x$ and $Q_y$  coordinates can be equivalently represented in polar vibrational coordinates $\rho$ and  $\phi$. Then, instead of quantum numbers $v_x$ and $v_y$, it is convenient to define $v_2=v_x+v_y$ and $l_2$, where $l_2$ is associated with the motion along $\phi$ coordinate: vibrational angular motion  around the symmetry axis. Thus, the vibrational energy is determined only by the quantum numbers $v_1$ and $v_2$: $E=\omega_1(v_1+1/2)+\omega_2(v_2+1)$. The number $v_2$ shows how many vibrational quanta are in the asymmetric mode. The number $l_2$ determines how many of the asymmetric quanta $v_2$  contribute to the vibrational angular momentum, $-v_2\le l_2\le v_2$ . In reality, due to the anharmonicity of potentials, the vibrational energy of states with same $v_1$ and $v_2$ but different $l_2$ are slightly different. However, pairs of states with $\pm l_2$, where $l_2\ne 3\tilde k$ (here and below, $\tilde k$ is any integer number), are  strictly degenerate. This is a consequence of the fact that the $D_{3h}$ symmetry group has doubly degenerate representations. Thus, vibrational wave functions,
 \begin{equation}
\Phi_{\mathbf v}({\cal Q})=|v_1,v_2^{l_2}\rangle,
\end{equation}
of the ion are specified by the triad of  quantum numbers $v_1,v_2^{l_2}$.  The quantum number $l_2$ can have values $-v_2,-v_2+2 \dots v_2-2,v_2$ and it controls the symmetry of the vibrational wave functions. States with   $l_2=3\tilde k$, with an integer $\tilde k\ne 0$, can be of $A_1$ or $A_2$ symmetry. In order to distinguish the two symmetries using the number $l_2$, we will label states  $A_1$ with positive $l_2$, $l_2=3\tilde k,\ \tilde k\ge 0$ and $A_2$ states with negative $l_2$. A pair of states with $l_2\ne3\tilde k$ having both signs of $l_2$ constitutes the degenerate pair of functions, that transform according to the $E$ representation. In contrast to the numbers $v_1,v_2^{l_2}$, the classifications with symmetry labels $A_1$, $A_2$, or $E$ are exact.

In the present treatment the relative phase of degenerate states $|v_1,v_2^{l_2}\rangle$ with $l_2\ne 3\tilde k$ is slightly different from the one used in Ref. \cite{watson00}. Transformations of $|v_1,v_2^{l_2}\rangle$ with the present choice of phases are summarized in Table III of Ref. \cite{kokoouline03b}. 

The third factor of the total ionic wave function is the nuclear spin wave function. The nuclear spin states are classified according to the total spin $I=3/2$ or $I=1/2$. These states are constructed as described in Ref. \cite{kokoouline03b}. The result is   $\Phi^I_0(I=3/2)$, $\Phi^I_{-1}(I=1/2)$, and $\Phi^I_{+1}(I=1/2) $ wave functions, transformed according to the $A_1\oplus E$ representation of the respective symmetry group $S_3$ of three identical particle permutations. The state $\Phi^I_0(M=3/2)$  transforms according to the $A_1$ representation; the states $\Phi^I_{-1}(I=1/2)$ and $\Phi^I_{+1}(I=1/2)$ transform according to the $E$ representation.
 
The transformation of the total wave function  $\Phi_{t}^{n.sym}$ is determined by the quantum numbers $K^+,l_2$, and $g_I$ . The final step in the construction of the total wave function is an appropriate symmetrization of $\Phi_{t}^{n.sym}$. Since the total wave function should be antisymmetric with respect to (12) we determine $\Phi_{t}$:
\begin{eqnarray}
\label{eq:final_tot_wf}
\Phi_{t}=\frac{1}{\sqrt{2}}\left(\Phi_{t}^{n.sym}(K^+,l_2,g_I)-(-1)^{N^+}s_2\Phi_{t}^{n.sym}(-K^+,l_2',-g_I)\right),
\end{eqnarray}
In the above equation $s_2=1$ for all vibrational states excluding $A_2$, for which $s_2=-1$; $l_2=-l_2'$ if $l_2= 3\tilde k\pm 1$, and $l_2=l_2'$ if $l_2= 3\tilde k$. The condition for antisymmetry with respect to (12) is specified explicitly. If the symmetrization is trivial, i.e. both terms in Eq. (\ref{eq:final_tot_wf}) are identical, then the wave function is
\begin{eqnarray}
\label{eq:final_tot_wf_triv}
\Phi_{t}=\Phi_{t}^{n.sym}(K^+,l_2,g_I).
\end{eqnarray}
It is only possible if $g_I=0$, $K^+=0$ and $l_2=l_2'$.

The fermionic nature of nuclei also requires that the wave function  $\Phi_{t}$ should be  antisymmetric with respect to operations of (13) and (23). It is only possible if $\Phi_{t}$ transforms according to the  $A_2'$ or $A_2''$ representations of the $D_{3h}(M)$ group. This condition can be written as  $\tilde G=3\tilde k$, where $\tilde G=K^++l_2+g_I$. The determination of the total symmetry has one exception from the above rule. Namely, when the symmetrization is trivial: $l_2=-l_2',\ K^+=0$. For this case, the rovibrational part of the product (\ref{eq:final_tot_wf_triv}) has $A_1$ or $A_2$ symmetry, thus, $g_I$ can  only be 0.  Finally, the overall parity of the total state, which is determined as transformational under the operation of total inversion $E^*$, is determined by the number $K^+$: The parity is even if $K^+$ is even and the parity is odd if $K^+$ is odd.

In this study we are primarily interested in the ortho-modification of the H$_3$ molecule of $A_2'$ symmetry. Thus, Eq. (\ref{eq:final_tot_wf}) is reduced to
\begin{eqnarray}
\label{eq:rv_tot_wf_ortho}
\Phi_{t}=\frac{1}{\sqrt{2}}\left({\cal R}_{N^+K^+m^+}(\alpha , \beta , \gamma)|v_1,v_2^{l_2}\rangle-(-1)^{N^+}s_2|{\cal R}_{N^+ -K^+m^+}(\alpha , \beta , \gamma)|v_1,v_2^{l_2'}\rangle\right)
\end{eqnarray}
or, when $K^+=0$ and $l_2=l_2'$, to
\begin{eqnarray}
\label{eq:rv_tot_wf_ortho_triv}
\Phi_{t}={\cal R}_{N^+K^+m^+}(\alpha , \beta , \gamma)|v_1,v_2^{l_2}\rangle.
\end{eqnarray}

Only states with $\tilde G= K^++l_2=3\tilde k$ and with even $K^+$ are allowed. Again there is an exception, when the symmetrization is trivial: When $K^+=0$ and $N^+$ is even (rotational symmetry is $A_1'$), $l_2=3\tilde k$ can only be negative ($A_2$ vibrational symmetry); when  $K^+=0$ and $N^+$ is odd, $l_2=3\tilde k$ must be positive or zero.

\subsection{ H$_3^+$ vibrational dynamics in an adiabatic hyperspherical approach}
\label{sec:DR_in_HSC}

As in our study of H$_3^+$ dissociative recombination \cite{kokoouline03a,kokoouline03b,kokoouline01}, we employ the adiabatic hyperspherical approach to describe the vibrational dynamics of H$_3^+$ and  H$_3$ in three dimensions. In this approach, three hyperspherical coordinates, the hyper-radius $R$ and two hyperangles  $\theta$, $\varphi$, represent three vibrational degrees of freedom \cite{kokoouline03b,zhou93,lin95,esry96}. In our calculations we use accurate potential  surfaces of H$_3^+$ from Refs. \cite{cencek98,jaquet98}.

\begin{figure}[h]
\includegraphics[height=17cm]{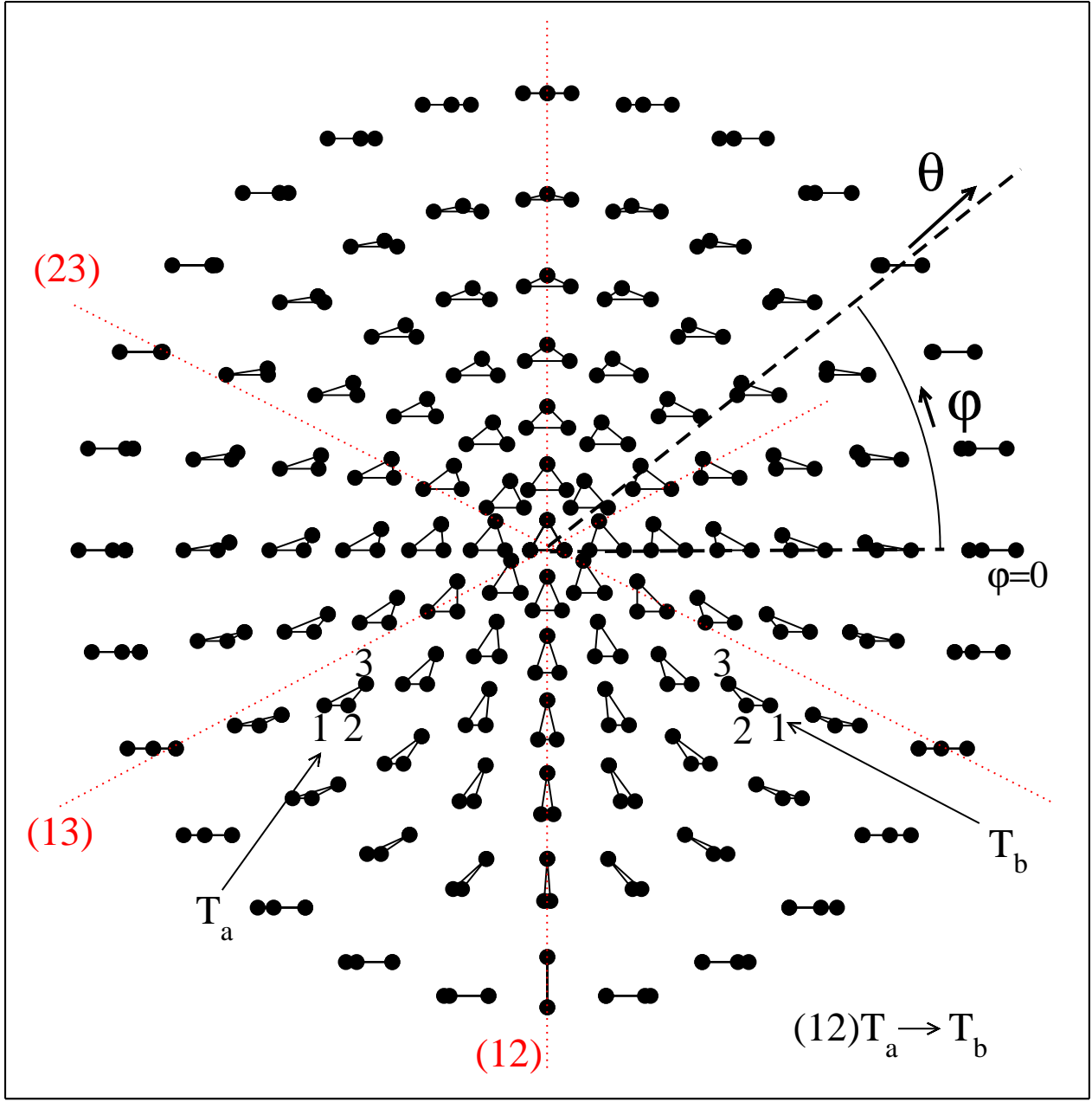}
\caption{\label{fig:radius_angles}
The figure demonstrates how the two-dimensional space of the hyperangles $\theta$ and $\varphi$ reproduces all possible shapes of a triangle consisting of three nuclei. The size of the system is controlled by the hyper-radius, which is not shown here. Hyperangle $\varphi$ starts at point  $\varphi=0$ and goes counter clockwise from 0 to $2\pi$. The hyperangle $\theta$ is 0 at the symmetric configuration and it changes to $\theta=\pi/2$ where it represents linear configurations. The figure also shows  symmetry axes: Every operation in the group $C_{3v}$ involves only the hyperangle $\varphi$. For example, the operation (12) transforms the triangle  $T_a$ into the triangle  $T_b$, which corresponds to a reflection around the line $\varphi=\pi/2\dots 3\pi/2$.
}
\end{figure}

The hyperspherical coordinates used are symmetry-adapted coordinates: Each operation of the $C_{3v}$ group--a permutation of instantaneous positions of the three nuclei--is described with an appropriate change in the hyperangle $\varphi$ only. With this respect, the hyperspherical coordinates are similar to the normal coordinates of H$_3^+$ \cite{mistrik00,kokoouline03b,stephens95}, where all operations from $C_{3v}$ involve the polar angle $\phi$ uniquely. For example, the effect of the $(123)$ operation is a cyclic permutation of the three internuclear distances. In the hyperspherical coordinates, it is realized by adding the angle $2\pi/3$ to $\varphi$ as determined in Eqs. (23) of Ref. \cite{kokoouline03b}. The operation $(12)$ exchanges the internuclear distances $r_2$ and $r_3$. Equations (23) of Ref. \cite{kokoouline03b} show that this operation corresponds to a mirror reflection about the axis $\varphi_{(12)}=[-\pi/2,\pi/2]$. Only the angle $\varphi$ is changed, into $\varphi '=(12)\varphi=\pi-\varphi$. This operation is exhibited in Fig. \ref{fig:radius_angles}: The operation $(12)$ exchanges  nuclei 1 and 2, transforming the triangle $T_a$ into $T_b$. The figure also shows all three symmetry axes, corresponding to three binary permutations $(12)$, $(23)$, and $(13)$. These symmetry properties of the hyperspherical vibrational coordinates simplify our treatment appreciably.

In the adiabatic hyperspherical method we first solve the vibrational Schr\"odinger equation at a fixed hyper-radius $R$ \cite{kokoouline03b},  obtaining a set of energies $U_i$ and corresponding eigenfunctions $\Phi_i(\theta,\varphi)$. Changing $R$, we obtain a set of adiabatic potential curves $U_i(R)$ and adiabatic hyperspherical eigenstates $\Phi_i(\theta,\varphi; R)$. As mentioned above, each element from the $C_{3v}$ symmetry group is represented by a corresponding transformation involving only the  hyperangle $\varphi$. The hyper-radius is not involved in the  $C_{3v}$ operations. Thus, the vibrational hyperspherical states $\Phi _{i }(\theta ,\varphi; R)$ and curves $U_i(R)$ can be classified according to irreducible representations of the group $C_{3v}$. Namely, each state $\Phi _{i}(\theta ,\varphi; R)$ and corresponding curve $U_i(R)$ can be labeled by either the $A_1$, $A_2$, or $E$ irreducible representation. The representation $E$  is two-dimensional. Thus, two degenerate $E$-components  $\Phi _{i}(\theta ,\varphi)$ will be labeled by $E_a$ and $E_b$. Their linear combinations are also good vibrational eigenstates.

\begin{figure}[h]
\includegraphics[width=17cm]{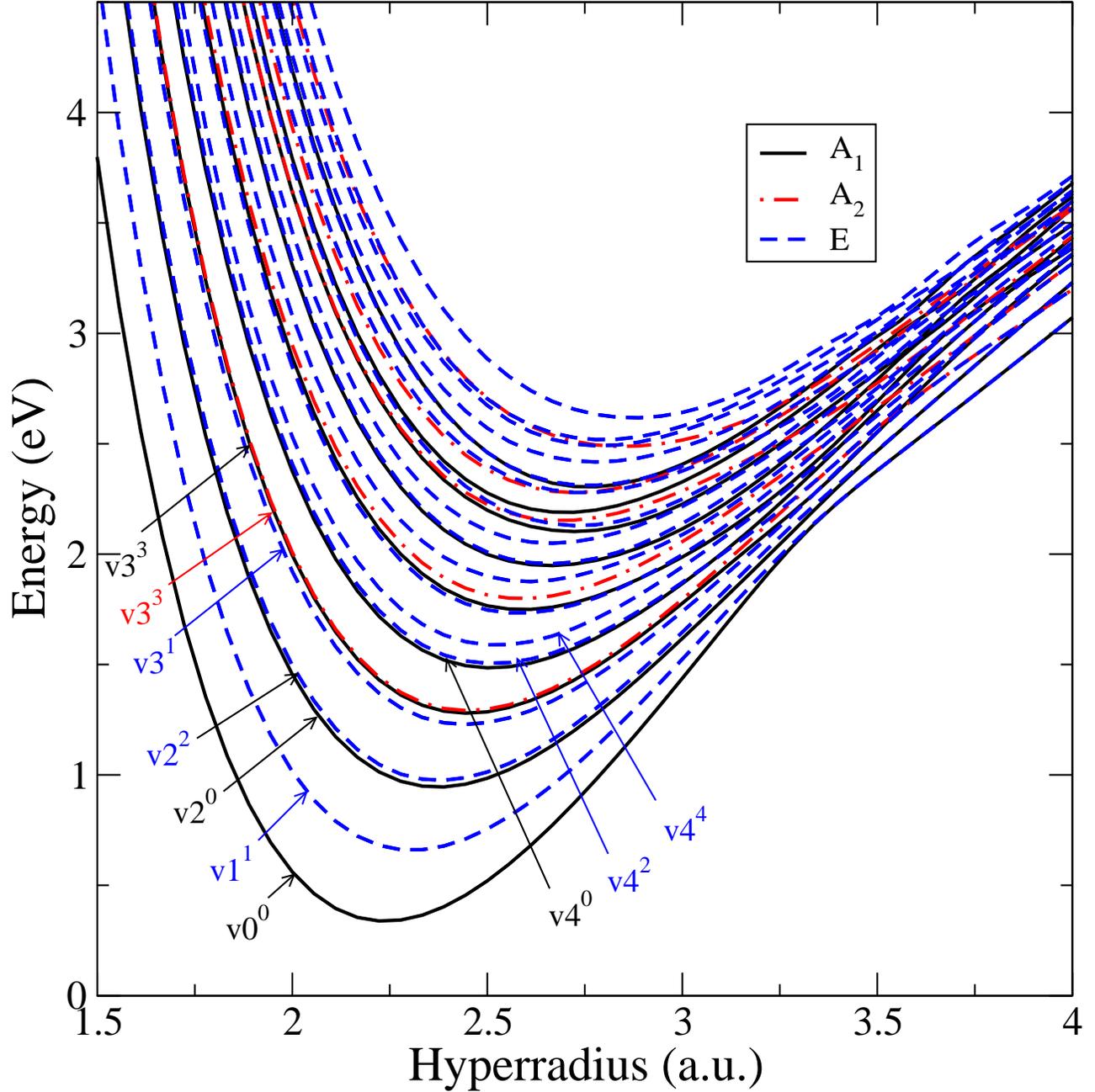}
\caption{\label{fig:Ion_potentials}
Lowest adiabatic hyperspherical potentials of the H$_3^+$ ion plotted as functions of hyper-radius. Potentials of the $A_1$ symmetry are plotted with solid lines, $A_2$ curves are represented with dot-dashed lines, and $E$ curves are represented with dashed lines. Every $E$ curve is doubly degenerate. For several lowest curves we also specify approximate  quantum numbers $v_2$ and $l_2$. For excited hyperspherical states such quantum numbers can not be defined since the three-dimensional ionic potential is strongly anharmonic at large energies. As is clear in the figure the potential $v_14^4$ is already strongly shifted from its partners  $v_14^2$ and $v_14^0$ toward to the next family of states $v_15^{l_2}$.
}
\end{figure}

 Several low ionic hyperspherical curves $U_{i}^{+}(R)$ are shown in Fig. \ref{fig:Ion_potentials}. We specify the pair $v_2^{l_2}$ of quantum numbers for first several  states. As in the familiar Born-Oppenheimer approximation for diatomic molecules, curves of the same symmetry do not cross, whereas curves of different symmetries may cross. One can see from Fig. \ref{fig:Ion_potentials} that low-lying potential curves with the same $v_2$ are almost degenerate. This is because the anharmonicity is quite small for low states. However, at $v_2=4$ the potential curves with quantum numbers $v_2^{l_2}=4^4$ and $v_2^{l_2}=4^{0,2}$ are already significantly separated.

Similar to Ref. \cite{kokoouline03b}, from real-valued $E$-states obtained after a diagonalization at fixed $R$, in the space of the two hyperangles $\theta$ and $\varphi$, we construct ``helicity'' \cite{varshalovich} $E$ states:
\begin{eqnarray}
|v_2^{l_2}\rangle= \frac{1}{\sqrt{2}} (E_a+ iE_b); \nonumber\\
|v_2^{l_2'}\rangle=|v_2^{-l_2}\rangle=\frac{1}{\sqrt{2}} (E_a- iE_b).
\end{eqnarray}
The sign of $l_2$ in the above equation is chosen in such a way that (123) transforms the state $|v_2^{\pm l_2}\rangle$ as 
\begin{eqnarray}
\label{eq:E_rot}
(123)|v_2^{\pm l_2}\rangle=e^{\frac{\pm 2\pi i l_2}{3}}|v_2^{\pm l_2}\rangle.
\end{eqnarray}
For example, if $|l_2|=2$, then $l_2=-2$ and $l_2'=2$ (see discussion in Sec. \ref{subsec:sym_comp}). Finally, we multiply all real-valued vibrational functions $A_2$  by $i$ in order to obtain a real reaction matrix $K$.

Once the adiabatic hyperspherical potential curves $U_{i}^{+}(R)$ are determined, we calculate vibrational energies, $E_{i,v}$, by solving the adiabatic hyper-radial equation,
\begin{eqnarray}
\label{eq:schr_hyperradius}
\left[-\frac{1}{2\mu}\frac{\partial^2}{\partial R^2}+U_{i}^{+}(R)\right]\psi^+_{i,v}(R)=E_{i,v}\psi^+_{i,v}(R),
\end{eqnarray}
where $\{v,i\}\equiv\{v_1,v_2^{l_2}\}$. In solution of Eq. (\ref{eq:schr_hyperradius}), we seek solutions (Siegert states) that obey outgoing wave boundary conditions \cite{tolstikhin97,tolstikhin98} at a finite hyper-radius $R_0$ and which are normalized as in Ref. \cite{kokoouline03b}.

Inclusion of Siegert states into the treatment allows us to represent the dissociation of the neutral H$_3$ formed during the collision between H$_3^+$ and the incident $e^-$.

\subsection{Comparison with an alternative symmetrization procedure}
\label{subsec:sym_comp}

An alternative procedure for the symmetrization of the total wave function was proposed by Spirko {\it et al.} \cite{spirko85}. Spirko {\it et al.} describe how rovibrational wave functions of different $D_{3h}$ representations, $\Gamma = A_1^\zeta,A_2^\zeta,E_a^\zeta$, and $E_b^\zeta$, are obtained from the products of rotational and vibrational parts. We use the symbol $\zeta$ to specify the parity of a state, where $\zeta='$ or  $\zeta=''$. (In Ref. \cite{spirko85}, the symbol $\dag$ was used for this purpose.)
In the approach of Ref.  \cite{spirko85}, the rotational states with  $\Gamma = A_1^\zeta,A_2^\zeta,E_a^\zeta$, or $E_b^\zeta$ are obtained by combining the symmetric top states $|N^+,K^+,m^+ \rangle$ (Eqs. (59)-(61) of Ref. \cite{spirko85}). Then  products of rotational and vibrational states are constructed. At the next step, these products are symmetrized again to give rovibrational states of good symmetry. This final step is quite laborious since one has to consider all possible combinations of different rotational and vibrational states (Eqs. (62)-(77) of Ref. \cite{spirko85}). To properly include all states of different nuclear spin symmetries, one would have to construct nuclear spin states and then use a similar symmetrization procedure (Eqs. (62)-(77) of Ref. \cite{spirko85}) one more time. In implementing the procedure of Spirko {\it et al.} for construction of the $e^-+$H$_3^+$ scattering matrix, we have found that it is difficult to obtain all states of the right symmetry. When carried out incorrectly, our scattering matrix displayed non-zero matrix elements between states of different symmetries, which of cause signals an error. For this reason, we think that our symmetrization procedure is advantageous for the present study, where we typically include and symmetrize hundreds of states. Our procedure involves only two simple steps and can be easily automated on a computer. First, the products of rotational, vibrational, and nuclear spin non-symmetrized states are constructed. For each product, the number $\tilde G$ [or the total symmetry for the states that are trivially symmetrized, such as Eq. (\ref{eq:rv_tot_wf_ortho_triv})] is determined. It describes behavior of the product under the symmetry operations  (12) and (123). Then the symmetrization procedure (if it is not trivial) is accomplished by a single equation [see Eq. (\ref{eq:final_tot_wf})].

To facilitate the comparison with the procedure, proposed by Spirko {\it et al.} \cite{spirko85}, we introduce a ``helicity'' pair of degenerate $E$ states, which transform in a uniform way, irrespective of their nature: rotational, vibrational, or nuclear spin. We determine two degenerate states $|E_+\rangle$ and $|E_-\rangle$ by their symmetry properties 
\begin{eqnarray}
\label{eq:Epm}
(123)|E_\pm\rangle=e^{\pm i\frac{2\pi}{3}} |E^\pm\rangle,  \nonumber\\
(12)|E_\pm\rangle=|E_\mp\rangle,
\end{eqnarray}
irrespective of coordinates: rotational, vibrational, electronic or nuclear spin. Using Table II of Ref. \cite{kokoouline03b}, it can be easily verified that rotational  $|E^{r,\zeta}_\pm\rangle$ states are obtained from $|N^+,K^+,m^+ \rangle$ according to
\begin{eqnarray}
\label{eq:Epmr}
|E^{r,\zeta}_+\rangle=|N^+, K^+,m^+ \rangle,\ \text{if } K^+=3\tilde k +1 \nonumber,\\
|E^{r,\zeta}_-\rangle=(-1)^{N^+}|N^+,-K^+,m^+ \rangle, \  \text{if } K^+=3\tilde k +1,\\
|E^{r,\zeta}_+\rangle=|N^+, -K^+,m^+ \rangle,\  \text{if } K^+=3\tilde k +2 \nonumber,\\
|E^{r,\zeta}_-\rangle=(-1)^{N^+}|N^+,K^+,m^+ \rangle,\  \text{if } K^+=3\tilde k +2.\nonumber
\end{eqnarray}
In the above equation, $\tilde k$ is a non-negative integer. Vibrational $|E^v_\pm\rangle$ states are obtained using Eq. (\ref{eq:E_rot}) as
\begin{eqnarray}
\label{eq:Epmv}
|E^v_\pm\rangle=|v_1,v_2^{\pm |l_2|}\rangle ,\ \text{if } |l_2|=3\tilde k +1, \nonumber\\
|E^v_\pm\rangle=|v_1,v_2^{\mp |l_2|}\rangle,\  \text{if } |l_2|=3\tilde k +2.
\end{eqnarray}

In addition to  $|E_\pm\rangle$ states we introduce another pair of states $|E_a\rangle$ and $|E_b\rangle$ as
\begin{eqnarray}
\label{eq:Eab1}
|E_\pm\rangle = \frac{1}{\sqrt{2}} (|E_a\rangle \pm i|E_b\rangle).
\end{eqnarray}
or
\begin{eqnarray}
\label{eq:Eab2}
|E_a\rangle = \frac{1}{\sqrt{2}} (|E_+\rangle + |E_-\rangle),\nonumber\\
|E_b\rangle = \frac{1}{i\sqrt{2}} (|E_+\rangle - |E_-\rangle).
\end{eqnarray}
The states $|E_a\rangle$ and  $|E_b\rangle$ are real. Using Eqs. (\ref{eq:Epm}) and (\ref{eq:Eab2}), we obtain that
\begin{eqnarray}
\label{eq:12Eab}
(12)|E_a\rangle = |E_a\rangle,\nonumber\\
(12)|E_b\rangle = -|E_b\rangle.
\end{eqnarray}
Using Eqs. (\ref{eq:Epm}), (\ref{eq:Eab1}), and (\ref{eq:Eab2}) and the fact that $e^{\pm i\frac{2\pi}{3}}=-\frac{1}{2}\pm i\frac{\sqrt{3}}{2}$, we have for the operation (123)
\begin{eqnarray}
\label{eq:123Eab}
(123)|E_a\rangle = -\frac{1}{2}|E_a\rangle-\frac{\sqrt{3}}{2}|E_b\rangle,\nonumber\\
(123)|E_b\rangle = \frac{\sqrt{3}}{2}|E_a\rangle-\frac{1}{2}|E_b\rangle.
\end{eqnarray}

At this stage, we can compare the present convention for $|E_a\rangle$ and $|E_b\rangle$ states with the one from Ref. \cite{spirko85}. Comparing formulas (\ref{eq:12Eab}) and (\ref{eq:123Eab}) with Eqs. (51)-(54) of Ref. \cite{spirko85}, we conclude that the two conventions for $|E_a\rangle$ and $|E_b\rangle$ vibrational states coincide. A comparison for  rotational $|E_a\rangle$ and $|E_b\rangle$ states should account for a different choice of coordinate axes made in \cite{spirko85} and in the present study. In order to compare with the present work, axes $x$ and $y$ in \cite{spirko85} must be exchanged. This will affect Eqs. (60)-(61) in \cite{spirko85}: The factor $(-1)^J$ in Eqs. (60)-(61) at the second term of the equations must be omitted. After this modification,   Eqs. (60) and (61) in \cite{spirko85} describe exactly the same states as the rotational states determined in Eqs. (\ref{eq:Epmr}) and (\ref{eq:Eab2}) of the present study.

Now we can derive formulas (62)-(77) of  \cite{spirko85} for the product of rotational and vibrational states. Consider, for example, Eq. (75) of \cite{spirko85} describing an overall $E_b$ state composed from $|E_a\rangle$ and $|E_b\rangle$ rotational and vibrational states. Using Eq. (\ref{eq:Eab2}) and the fact that $|E_+\rangle=|E_-\rangle\otimes|E_-\rangle$ and $|E_-\rangle=|E_+\rangle\otimes|E_+\rangle$, we derive
\begin{eqnarray}
\label{eq:75}
|E_b^{\zeta}\rangle=\frac{1}{i\sqrt{2}}\left(|E_+\rangle-|E_-\rangle\right)=\frac{1}{i\sqrt{2}}\left(|E^v_-\rangle|E^{r,\zeta}_-\rangle - |E^v_+\rangle|E^{r,\zeta}_+\rangle\right)=\nonumber\\
\frac{1}{i2\sqrt{2}} \left[\left(|E^v_a\rangle - i |E^v_b\rangle\right)\left(|E^{r,\zeta}_a\rangle - i |E^{r,\zeta}_b\rangle\right)-\left(|E^v_a\rangle + i |E^v_b\rangle\right)\left(|E^{r,\zeta}_a\rangle + i |E^{r,\zeta}_b\rangle\right)\right]=\nonumber\\
-\frac{1}{i\sqrt{2}}\left(|E^v_a\rangle |E^{r,\zeta}_b\rangle + |E^v_b\rangle |E^{r,\zeta}_a\rangle\right).
\end{eqnarray}
Comparing with Eq. (75)  of  \cite{spirko85} we see that the $|E_b\rangle$ function of Eq. (75) differs from our state by an overall sign. The sign is important since it affects the result of the (123) operation. (The corresponding $|E_a\rangle$ state (Eq. (70) \cite{spirko85}) has the correct sign.) Note that Eq. (75) of \cite{spirko85}) also has a typographical error: Instead of symbol $A_2$ in the first term, there should be $|E_a\rangle$.

The rest of the formulas, (62)-(77) in \cite{spirko85}, can be derived in a similar way.

\section{The scattering matrix for en electron colliding with H$_3^+$}
\label{sec:S}

\subsection{Short-range scattering matrix of H$_3^+ +e^-$ in presence of Siegert vibrational pseudostates}

Once vibrational Siegert pseudostates are calculated, the scattering matrix, describing the collision of the electron with the vibrating  H$_3^+$ ion, can be constructed. A vibrational frame transformation \cite{greene85} can be used to calculate the amplitude $S_{i,i'}$ for the scattering from one vibrational state $i'$ to another $i$.  Here, the indices $i$ and $i'$ enumerate vibrational states and states of different  projections $\Lambda$ of the electron angular momentum. The vibrational part of the indices is represented by the triad $\{v_1,v_2^{l_2}\}$: The pair $v_2^{l_2}$ represents one hyperspherical curve $U_i^+(R)$; the index $v_1$ enumerates the Siegert pseudostates lying within that curve. Therefore the amplitude $S_{i,i'}$ for the process
\begin{equation}
\label{eq:S_v}
e^-(\Lambda')+\text{H}_3^+(v_1',v_2'^{l_2'})\to e^-(\Lambda)+\text{H}_3^+(v_1,v_2^{l_2})
\end{equation}
is calculated in two steps, similar to the two-step calculation of vibrational energies. First, we determine $R$-dependent amplitude
\begin{eqnarray}
\label{eq:S_hangles}
S_{v_2,l_2,\Lambda;v_2',l_2',\Lambda'}(R)=\Big<\big<\Phi _{v_2,l_2 }\Big|S_{\Lambda;\Lambda'}({\cal Q})\Big|\Phi _{v_2',l_2 '}\big>\Big>_{(\theta,\varphi)},
\end{eqnarray}
where the double brackets means an integration over hyperangles at constant hyper-radius $R$, and $\cal Q$ represents three internuclear coordinates. The scattering matrix $S_{\Lambda;\Lambda'}({\cal Q})$ includes the Jahn-Teller interaction and is calculated from the reaction matrix $K$ as described in Ref. \cite{kokoouline03b} (see Eqs. (18)-(20) of Ref. \cite{kokoouline03b}).

The scattering matrix $S({\cal Q})$ in Eq. (\ref{eq:S_hangles}) has indices $\Lambda$ and $\Lambda'$ and represents an amplitude for the process:
\begin{equation}
\label{eq:S1}
e^-(l=1,\Lambda')+\text{H}_3^+({\cal Q})\to e^-(l=1,\Lambda)+\text{H}_3^+({\cal Q}).
\end{equation}
Therefore,  $S_{\Lambda, \Lambda'}({\cal Q})$ represents the scattering amplitude when the electron scatters from one channel $\Lambda$ to another $\Lambda'$, while the nuclei do not have time to move. Equation (\ref{eq:S1}) describes the short-range H$_3^+ + e^-$ collision in the clumped-nucleus approximation, where nuclear degrees of freedom are not yet coupled to the electronic degrees of freedom. 

The equation for the second step reads  similarly as:
\begin{eqnarray}
\label{eq:S_hradius}
S_{v_1,v_2,l_2,\Lambda;v_1',v_2',l_2',\Lambda'}=\big<\psi_{v_1,v_2,l_2 }(R)\big|S_{v_2,l_2,\Lambda;v_2',l_2',\Lambda'}(R)\big|\psi_{v_1',v_2',l_2 '}(R)\big>_{S},
\end{eqnarray}
where brackets $\big<\big>_{S}$ means the integration in the sense of Siegert pseudostates, i.e. with an implied surface term \cite{tolstikhin97,tolstikhin98,hamilton02,kokoouline03b}:

\begin{eqnarray}
\label{eq:int_Sieg}
\big<\psi_{v_1,v_2,l_2}(R)\big|S_{v_2,l_2,\Lambda;v_2',l_2',\Lambda'}(R)\big|\psi_{v_1',v_2',l_2 '}(R)\big>_{S}=\nonumber\\
\int_0^{R_f}\psi_{v_1,v_2,l_2 }(R)\cdot S_{v_2,l_2,\Lambda;v_2',l_2',\Lambda'}(R)\cdot \psi_{v_1',v_2',l_2 '}(R)dR+\\
i\frac{\psi_{v_1,v_2,l_2}(R_f)S_{v_2,l_2,\Lambda;v_2',l_2',\Lambda'}(R_f)\psi_{v_1',v_2',l_2'}(R_f)}{k_{v_1,v_2,l_2}+k_{v_1',v_2',l_2'}}.\nonumber
\end{eqnarray}
When the integral in the above equation is evaluated, the usual complex conjugation of the bra wave function $\psi_{v_1,v_2,l_2}(R)$ is omitted. The quantity $k_{v_1,v_2,l_2}$ is a complex wave number obtained from the complex energy $E_{v_1,v_2,l_2}$ of the corresponding Siegert state $|v_1,v_2^{l_2}\rangle$  \cite{hamilton02,kokoouline03b}:
\begin{equation}
E_{\{v_1,v_2,l_2\}}=k^2_{\{v_1,v_2,l_2\}}/(2\mu)+D_{v_2,l_2}
\end{equation}
where $D_{v_2,l_2}$ is the dissociation limit of the corresponding adiabatic hyperspherical curve $U_{v_2,l_2}(R)$. In the present approach $D_{v_2,l_2}$ is approximated by a value of $U_{v_2,l_2}(R)$ at large hyper-radius $R_f$,  $D_{v_2,l_2}=U^+_{v_2,l_2}(R_f)$.

 Due to the presence of Siegert states with complex eigenenergies, this electron-ion scattering matrix is not unitary. The non-unitarity accounts for the fact that the electron can become stuck in the ion, leading to the dissociation of the system into neutral products.

\subsection{Rotational frame transformation and the final short-range scattering matrix}

The electron-ion scattering matrix  $S_{i,i'}$, constructed above, does not account for the possibility of rotational excitation of the ion. If the H$_3^+$ ion is initially in one rotational state $(N^{+'},K^{+'})$, a collision with the electron can scatter the rotational state into $(N^+,K^+)$. Thus, an element ${\cal S}_{i,i'}$ of the total scattering matrix ${\cal S}$ describes a transition from one rovibrational state $i'=\{v_1',v_2'^{l_2'}\}(N^{+'},K^{+'})$ to another $i=\{v_1,v_2^{l_2}\}(N^+,K^+)$.   In indices $i$ and $i'$, we do not specify quantum numbers that are conserved during the collision. These quantum numbers are the total energy $E$, the total nuclear spin $I$ of H$_3^+$, the total angular momentum of the system $N$ and its projection  $m$  on the laboratory $z$-axis. (see Eq. (\ref{eq:trans2}) below) and, finally the total symmetry of the system: $A_2'$ or $A_2''$.

The change in the rotational excitation $(N^{+'},K^{+'})\to (N^+,K^+)$ is taken into account using the  rotational frame transformation approximation \cite{greene85,jungen96}: Such a transition occurs mainly when the electron approaches close to the ion. Since a basis of rotational functions exists for which the short-range rotational Hamiltonian is diagonal, the transition amplitude for $(N^{+'},K^{+'})\to (N^+,K^+)$ can be described by considering the coefficients that link the long-range quantum numbers $(N^+,K^+)$ with the short-range quanta. The short-range rotational states are specified by the projection $\Lambda$  of the electronic angular momentum $l$ on the ion-fixed $Z$-axis  and by the projection $K$ of the total angular momentum $\vec N=\vec N^++\vec l$ of the neutral molecule on the same $Z$-axis. $N$ and $l$ are conserved quantum numbers in both rotational bases. (This is one approximation of our treatment, because in reality $l$-changing collisions can occur with a small amplitude.) Below, we present a detailed description of the rovibrational frame transformation for the $e^-+$H$_3^+$ system and specify all quantum numbers in both regions of interaction between $e^-$ and H$_3^+$.

At large electron-ion distances, the system is described by the electronic angular momentum $l$ and its projection $\lambda$ on the laboratory $z$-axis, by the total ionic angular momentum $N$, its projection $m^+$ on the laboratory $z$-axis and its projection $K^+$ on the molecular symmetry axis $Z$. Correspondingly, we represent the wave function of the $e^-+$H$_3^+$ system by a product of $\Phi_{t}^{n.sym}$ and $Y_{l\lambda}(\theta,\varphi)$ (at this stage we do not specify electronic radial part of the total wave function):
\begin{equation}
\label{eq:wf_long}
{\cal R}_{N^+m^+K^+}(\alpha  \beta  \gamma)Y_{l\lambda}(\theta,\varphi){\cal V }(v_1,v_2^{l_2})\Phi_I .
\end{equation}
The angles $\theta,\varphi$ are spherical angular coordinates of the electron in the laboratory coordinate system (LS). 

At short distances, the most appropriate molecular states, i.e., the states that almost diagonalize the Hamiltonian, are specified by the  projection $\Lambda$ of $\vec l$ on the molecular $Z$ axis; by three internuclear coordinates $\cal Q$ since these remain approximately frozen during a single collision; by the total angular momentum of the system $N$, including the electron momentum;  and by its projections on the  molecular $Z$ axis, $K$, and on the laboratory $z$-axis, $m$. Thus, the total wave function at short distances is
\begin{eqnarray}
\label{eq:wf_short2}
\big|N,K,m,\Lambda;v_1,v_2^{l_2}\rangle=(-1)^{l-\Lambda}{\cal R}_{N^+m^+K^+}(\alpha  \beta  \gamma) Y_{l\Lambda}(\theta',\varphi')|{\cal Q}\rangle \Phi_I.
\end{eqnarray}
The angles $\theta',\varphi'$ determine the position of the electron relative to the molecular coordinate system (MS). The transformation between the two wave functions is \cite{kokoouline03b}
\begin{equation}
\label{eq:trans2}
\big|N^+,K^+;N,m;v_1,v_2^{l_2}\rangle =\sum_\Lambda C_{l,-\Lambda;N,K}^{N^+,K^+}\big|N,K,m,\Lambda;v_1,v_2^{l_2}\rangle,
\end{equation}
which can be considered as if two angular momenta $\vec N$ and $\vec l$ with projections $K$ and $-\Lambda$ were added to give the momentum $\vec N^+$ with the projection $K^+=K-\Lambda$. The quantum numbers $N,l,m$ are not changed by the rotational, Eq.(\ref{eq:trans2}), nor vibrational frame transformations, Eqs. (\ref{eq:S_hangles} and (\ref{eq:S_hradius})). Therefore, they are good quantum numbers at short and long distances, within the approximation of this study.

Note that all three projections $\Lambda,K,K^+$ can be negative or positive. Equation (\ref{eq:trans2}) differs, for example, from the one in Ref. \cite{chang72} where all rotational functions are symmetrized with respect to different signs of projections. We keep both negative and positive projections explicitly in order to symmetrize  products of electronic, rotational, vibrational, and nuclear spin components of the total wave functions  at the very final step. As was mentioned above, this simplifies the symmetrization procedure.

The total short-range scattering matrix can now be constructed using the frame transformation techniques: When the electron is far from the ion,  the interaction Hamiltonian is diagonal in the basis of the long-range wave functions; at short distances, short-range wave functions almost diagonalize the Hamiltonian. The short-range Hamiltonian is not exactly diagonal in the basis of states of Eq. (\ref{eq:wf_short2}). It has off-diagonal elements in $\Lambda$, owing to the Jahn-Teller coupling. The following selection rules can be formulated: {\it (i)} The Hamiltonian can only couple vibrational states of the same vibrational symmetry and the same value of $\Lambda$, or {\it (ii)} A $p$-wave electron can couple the rovibrational channels according to the rule $(\Lambda=1,l_2=-1) \leftrightarrow (\Lambda'=-1,l_2'=1)$. These  selection rules insure that the total symmetry $\Gamma$ of the system is conserved during the collision.

The actual form of the coupling matrix $S_{\Lambda,\Lambda '}({\cal Q})$ is given in Refs.  \cite{stephens95,staib90a,staib90b}. The final scattering matrix is represented as
\begin{widetext}
\begin{eqnarray}
\label{eq:S_frame_trans}
{\cal S}_{N^+,K^+,v_1,v_2^{l_2};N^{+'},K^{+'},v_1',v_2'^{l_2'}}^{(N,K,m,l,I,\Gamma)}=\nonumber\\
= \sum_{\Lambda,\Lambda '} C_{l,-\Lambda ';N,K}^{N^{+'},K^{+'}}\left[\int_S{\cal V}(v_1',v_2'^{l_2'};{\cal Q})S_{\Lambda,\Lambda '}({\cal Q}){\cal V}(v_1,v_2^{l_2};{\cal Q})d{\cal Q}\right]C_{l,-\Lambda;N,K}^{N^+,K^+}.
\end{eqnarray}
\end{widetext}
The integral $\int_S$ in the above equation is evaluated according to Eqs. (\ref{eq:S_hangles}) and (\ref{eq:S_hradius}). The scattering matrix of Eq. (\ref{eq:S_frame_trans}) is diagonal over quantum numbers $N,K,l,m,I$ and $\Gamma$. In the above equation $\Gamma$ specifies the total molecular symmetry $\Gamma=A_2'$ or $A_2''$ of the considered state; $I$ specifies the total spin  1/2 or $3/2$.  Therefore, photoionization oscillator strengths can be calculated separately for all possible values of these quantum numbers. 

In practice, we calculate ${\cal S}$  using non-symmetrized states of the type  (\ref{eq:phi_total}) and (\ref{eq:wf_long}).  The symmetrization procedure of Eq. (\ref{eq:rv_tot_wf_ortho}) is then performed directly  on the scattering matrix. Let the total dimension of the matrix ${\cal S}_{N^+,K^+,v_1,v_2^{l_2};N^{+'},K^{+'},v_1',v_2'^{l_2'}}$ be $N_{tot}\times N_{tot}$.  For the full specification of the scattering process, the photoionization threshold energies of rovibrational states $i=\{v_1,v_2^{l_2}\}(N^{+},K^{+});\ i=1,2,\cdots N_{tot}$ are needed. We use accurate energies available in the literature \cite{lindsay01,mccall01}. For some excited rovibrational levels, however, where no data exist,  the energies were calculated using the adiabatic hyperspherical and rigid-rotor approximations. Another group of even higher energies are found to be complex, as expected for our Siegert pseudostate representation.

Equation (\ref{eq:S_frame_trans}) gives the $N_{tot}\times N_{tot}$ scattering matrix describing $e^- + $H$_3^+$ collisions. This matrix will be used to calculate the photoionization oscillator strengths.

\section{Oscillator strengths for the interpretation of H$_3$ photoionization experiments}
\label{sec:oscil_strength}

We apply our method to describe two photoionization experiments involving the H$_3$ molecule \cite{bordas91,mistrik00} . Both experiments have previously been successfully interpreted using multi-channel quantum defect theory \cite{mistrik00,stephens94,stephens95}. The present approach differs from  previous theoretical studies. Two main differences  are (1) the adiabatic hyperspherical method employed for vibrational degrees of freedom and (2) an inclusion of the previously missed factor $(-\pi)$ in the Jahn-Teller parameters $\delta$ and $\lambda$. Below we give a detailed description of how the dipole transition moments and oscillator strengths are calculated.

\subsection{Vibrational wave function of the initial state}

 In the experiment by Bordas {\it et al.} \cite{bordas91}, the  spectrum of the  photoionization process,
\begin{equation}
\label{eq:exp1}
\text{H}_3\ 3s\ A_2'\ (N'=1,K'=0, \{00^0\}) +  \omega_1 \to \text{H}_3^+ + e^- ,
\end{equation}
was measured. Here, $ \omega_1$ is the photon energy. In the second experiment by Mistr\'\i k {\it et al.} \cite{mistrik00}, an similar process was investigated. Mistr\'\i k {\it et al.} investigated photoionization starting from a different initial vibrational state,
\begin{equation}
\label{eq:exp2}
\text{H}_3\ 3s\ A_2'\ (N'=1,K'=0, \{10^0\}) +  \omega_2 \to \text{H}_3^+ + e^-,
\end{equation}
where $ \omega_2$ is the photon energy.
In both processes indicated, the initial symmetry $A_2'$ refers to the total molecular symmetry. The total molecular spin is $I=3/2$ in both experiments. The total symmetry can be viewed as a direct product of the symmetry $A_2'$ of the H$_3^+$ ion with $N^+=1,K^+=0, I=3/2$ and the symmetry $A_1'$ of the $3s$ electron. The dipole moment operator transforms according to irreducible representation $A_1''$ of the $D_{3h}$ group, since the dipole operator is proportional to the spherical harmonic $Y_{1m}(\theta',\phi')$. As a result, the final state of the electron-ion complex must have $A_2''$ total symmetry. Since we only consider final electronic states with $l=1$ (the symmetry is $A_1''$, when the electron is at large distances from the ion), the final symmetry of the ion should be $A_2'$.

As in to the previous theoretical treatments \cite{mistrik00,stephens94,stephens95}, in order to evaluate the dipole transition moments from the initial states in Eqs. (\ref{eq:exp1}) and (\ref{eq:exp2}), we use the $3s$ molecular potential surface of H$_3$ calculated by Nager and Jungen \cite{nager82}, in a Coulomb approximation. Vibrational wave functions of the $\{00^0\}$ and $\{10^0\}$ initial states were calculated using the adiabatic hyperspherical approach as described above.

\subsection{Scattering wave function of $e^-+$H$_3^+$: all asymptotic channels are open}
In the previous section, the scattering matrix for electron-ion collisions have been presented. However, for determination of transition dipole moments we need to know not only the scattering matrix, but also wave functions of corresponding scattering states. Following general quantum defect theory \cite{seaton83,aymar96}, we start with a scattering wave function assuming that the electron energy is so large that all possible entrance channels are open. We use the same phase conventions for wave functions as in Ref. \cite{aymar96}. The wave function $\vec \Psi_{i'}$ having out-going wave in the channel $i$ only can be represented as a $N_{tot}$ component vector, where each component $\Psi_{ii'}$ with $i=1,\cdots , N_{tot}$ corresponds to the incoming wave in channel $i$:
\begin{eqnarray}
\label{eq:Psifunction}
\Psi_{ii'}=\Phi_{i}(\omega)\frac{1}{i\sqrt{2}}\left(f_{i}^+(r)\delta_{ii'}-f_{i}^-(r){\cal S}^\dag_{ii'}\right),\nonumber\\
i=1,\cdots , N_{tot}.
\end{eqnarray}
The functions $f_{i}^\pm$ are outgoing/incoming waves in channel $i$. They are defined in Ref. \cite{aymar96}. The factor $\Phi_{i}(\omega)$ is a part of the total wave function; $\omega$ includes all degrees of freedom  excluding the radial one, $r$. The wave function (\ref{eq:Psifunction}) can be considered  as a complex conjugation of (or time-reversed to) the familiar incoming wave scattering state, having an incoming wave only in the channel $i'$. There are $N_{tot}$ functions of the type (\ref{eq:Psifunction}) and the whole set of $N_{tot}$ wave functions with $N_{tot}$ components can be considered as a $N_{tot}\times N_{tot}$ matrix $\mathbf \Psi$.

\subsection{Scattering wave function of $e^-+$H$_3^+$ when some channels are closed}

When the energy of the system is low enough such that some asymptotic channels are closed to ionization, the total wave function of the system must asymptotically (in the radial coordinate $r$) vanish in the corresponding channels. Thus, the total wave function differs from the one given by Eq. (\ref{eq:Psifunction}). Let $N_o$ and $N_c=N_{tot}-N_o$ represent the numbers of open and closed channels at a given total energy $E$.  In this situation: (i) there are only $N_o$ physically acceptable wave functions of the type (\ref{eq:Psifunction}) instead of $N_{tot}$;  (ii) these $N_o$ functions are zero at infinity in closed channels. Every acceptable wave function should have the following asymptotic behavior \cite{aymar96}:
\begin{eqnarray}
\label{eq:Psifunction_phys}
\Psi^{(-)}_{ii'}\to \Phi_{i}(\omega)\frac{1}{i\sqrt{2\pi k_i}}\left(e^{ik_ir}\delta_{ii'}-e^{-ik_ir}{\cal S}^{\dag phys}_{ii'}\right),\ i=1,\cdots , N_o; \nonumber\\
\Psi^{(-)}_{ii'}\to 0,\ i=N_o+1,\cdots , N_{tot}
\end{eqnarray} 
As is well-known, and shown in Refs. \cite{seaton83,aymar96}, one way to obtain  states with this asymptotic behavior is to construct linear combinations of states $\vec \Psi_{i'}$. In the matrix form this can be written as
\begin{eqnarray}
\label{eq:wf_phys}
\mathbf\Psi^{(-)}=\mathbf\Psi \mathbf B.
\end{eqnarray}
 The $N_{tot}\times N_o$ matrix $\mathbf\Psi^{(-)}$ consists of $N_o$ vectors, each having $N_{tot}$ components. The $N_{tot}\times N_o$ matrix $\mathbf B$ of the linear transformation is derived in Ref. \cite{aymar96}. If we partition the coefficient matrix into open and closed subspaces, as
\begin{eqnarray}
\mathbf B= \left( \begin{array}{c} \mathbf B_o  \\ 
\mathbf B_c  \end{array} \right),
\end{eqnarray}
the open-channel part $\mathbf B_o$ is represented by an $N_o\times N_o$ identity matrix and the closed part $\mathbf {B_c}$ is
\begin{eqnarray}
\mathbf {B_c} = -\left(\mathbf {S^\dagger_{cc}}-e^{2i\beta}\right)^{-1}\mathbf {S^\dagger_{co}}.
\end{eqnarray}
In the above equation the matrices $\mathbf {S^\dagger_{cc}}$ and $\mathbf {S^\dagger_{co}}$ are submatrices of  ${\cal S}^\dagger$ , which is itself partitioned as
\begin{eqnarray}
{\cal S}^{\dag } = \left( \begin{array}{cc} \mathbf{S^\dagger_{oo}} & \mathbf {S^\dagger_{oc}}  \\ 
\mathbf {S^\dagger_{co}} & \mathbf {S^\dagger_{cc}}  \end{array} \right)
\end{eqnarray}
and  $\beta(E)$ is a diagonal $N_c\times N_c$ matrix
\begin{equation}
\beta_{ij}(E)=\frac{\pi}{\sqrt{2(E_i-E)}}\delta_{ij},
\end{equation}
where $E_i$ refers to a particular ionization threshold $i=\{v_1,v_2^{l_2}\}(N^{+},K^{+})$.

After we apply the transformation (\ref{eq:wf_phys}), the component $\Psi^{(-)}_{oi'}$ of the $i$-th independent wave function $\vec \Psi^{(-)}_{i'}$ in the open channel $o$ is given outside the reaction volume by
\begin{eqnarray}
\label{eq:Psifunction_phys2}
\Psi^{(-)}_{oi'}= \Phi_{o}(\omega)\frac{1}{i\sqrt{2}}\left(f_o^+(r)\delta_{o,i'}-f_o^-(r) {\cal S}^{\dag phys}_{oi'}\right),\nonumber\\
o=1,\cdots , N_o,
\end{eqnarray}
where the physical $N_o\times N_o$ scattering matrix ${\cal S}^{\dag phys}$ is
\begin{eqnarray}
{\cal S}^{\dag phys} = \mathbf {S^\dagger_{oo}}- \mathbf {S^\dagger_{oc}}\left(\mathbf {S^\dagger_{cc}}-e^{2i\beta}\right)^{-1}\mathbf {S^\dagger_{co}}.
\end{eqnarray}
The closed-channel components $\Psi^{(-)}_{ci'}$ of $\vec \Psi^{(-)}_{i'}$ are determined by
\begin{eqnarray}
\Psi^{(-)}_{ci'}= \Phi_{c}(\omega)W_c(r)Z_{ci'},\nonumber\\
c=N_o+1 ,\cdots , N_{tot}.
\end{eqnarray}
In the above equation, $W_c(r)$ is the Whittaker function  and  $\mathbf Z$ is the $N_c\times N_o$ matrix of closed-channel coefficients (see Eqs. (2.52-2.54) of Ref. \cite{aymar96}). In a compact notation, the wave function $\vec \Psi^{(-)}_{i'}$ can be written as \cite{aymar96}:
\begin{eqnarray}
\label{eq:Psifunction_phys_full}
\vec\Psi^{(-)}= \sum_{o=1}^{N_o}\Phi_{o}(\omega)\frac{1}{i\sqrt{2}}\left(f_o^+(r)\delta_{o,i'}-f_o^-(r) {\cal S}^{\dag phys}_{oi'}\right)+\nonumber\\
\sum_{c=N_o+1}^{N_{tot}}\Phi_{c}(\omega)W_c(r)Z_{ci'}.
\end{eqnarray}

\subsection{Dipole moments of transitions from the initial bound state of H$_3$ to scattering states}

We need to evaluate $N_o$ dipole transition moments from a fixed initial state, $\Psi_{ini}=[3s^2A_2',\ \{00^0\}(1,0)]$ or $[3s^2A_2',\ \{10^0\}(1,0)]$ into all $N_o$ final states $\vec \Psi_f^{(-)}$. Each such moment is represented as 
\begin{eqnarray}
d_f=\langle\vec\Psi_f^{(-)}\big|\vec\epsilon\cdot \vec r\big|\Psi_{ini}\rangle,
\end{eqnarray}
where $\vec\epsilon$ is a unitary vector of laser light polarization. The initial state is closed for autoionization and, therefore, is represented as $\Psi_{ini}=\Phi_{ini}(\omega)W_{ini}(r)\nu_{ini}^{-3/2}$. The factor $\nu_{ini}^{-3/2}$ is due to the unity normalization of the initial bound state: The Whittaker function itself vanishes at infinity, but is chosen to have an energy-normalized amplitude at small $r$. Therefore, the dipole moment $d_f$ is written as
\begin{eqnarray}
\label{eq:dipole_long1}
d_f=\sum_{j=1}^{N_o}\langle  \Phi_{j}(\omega)\frac{1}{i\sqrt{2}}\left(f_j^+(r)\delta_{j,f}-f_j^-(r) {\cal S}^{\dag phys}_{jf}\right) \big|\vec\epsilon\cdot \vec r\big|\Phi_{ini}(\omega)W_{ini}(r)\nu_{ini}^{-3/2}\rangle +\nonumber\\
\sum_{j=N_o+1}^{N_{tot}}\langle  \Phi_{j}(\omega)W_j(r)Z_{jf}   \big|\vec\epsilon\cdot \vec r\big|\Phi_{ini}(\omega)W_{ini}(r)\nu_{ini}^{-3/2}\rangle .
\end{eqnarray}

The dipole moment $d_f$ above depends strongly on energy and, therefore, it must be calculated at a fine energy mesh. Calculation of all terms in Eq. \ref{eq:dipole_long1} is computationally expensive. However, inspecting the terms in  Eq. \ref{eq:dipole_long1}, we notice, that each term can be represented as a product of two factors: one factor depends strongly on energy, another factor is weakly energy-dependent. Briefly, the two sums above can be combined in one single sum of the form:
\begin{eqnarray}
\label{eq:dipole_long1a}
d_f= \sum_{j=0}^{N_{tot}}\tilde d_{j}\tilde v_{jf},
\end{eqnarray}
where each term of the sum is represented as product of two factors $\tilde d_{j}$ and $\tilde v_{jf}$:
\begin{eqnarray}
\label{eq:dipole_long2}
\tilde d_{j}=\left\{\begin{array}{ll}
\langle  \Phi_{f}(\omega)\frac{1}{i\sqrt{2}}f_f^+(r) \big|\vec\epsilon\cdot \vec r\big|\Phi_{ini}(\omega)W_{ini}(r)\nu_{ini}^{-3/2}\rangle, \text{ if } j=0\\
\langle  \Phi_{j}(\omega)\frac{1}{i\sqrt{2}}f_j^-(r) \big|\vec\epsilon\cdot \vec r\big|\Phi_{ini}(\omega)W_{ini}(r)\nu_{ini}^{-3/2}\rangle, \text{ if } j=1\dots N_o\\
\langle  \Phi_{j}(\omega)W_j(r)   \big|\vec\epsilon\cdot \vec r\big|\Phi_{ini}(\omega)W_{ini}(r)\nu_{ini}^{-3/2}\rangle,  \text{ if } j= N_o+1 \dots N_{tot}
\end{array} \right.
\end{eqnarray}

\begin{eqnarray}
\label{eq:dipole_long3}
\tilde v_{jf}=\left\{\begin{array}{ll}
1, \text{ if } j=0\\
 -{\cal S}^{\dag phys}_{jf}, \text{ if } j=1\dots N_o\\
Z_{jf} ,  \text{ if } j= N_o+1 \dots N_{tot}
\end{array} \right.
\end{eqnarray}
Notice that in Eq. (\ref{eq:dipole_long1a}) the summation starts at $j=0$ but not at $j=1$ as in Eq. (\ref{eq:dipole_long1}). This is because we represent the term with $j=f$ in (Eq. \ref{eq:dipole_long1}) as a sum of two terms in Eq. (\ref{eq:dipole_long1a}), with $j=0$ (corresponds to $f_j^+(r)\delta_{j,f}$) and  with $j=f$ (corresponds to $f_j^-(r) {\cal S}^{\dag phys}_{jf}$). This is necessary in order to separate energy dependent factors from energy independent ones. 

The partitioning given by Eq. (\ref{eq:dipole_long1a}) allows us to significantly reduce the calculation time, because over the small range considered in this calculation, it is adequate to evaluate $\tilde d_{j}$ only once for all energies.  However, $\tilde d_j$ should be calculated at every energy point.

\subsection{Transition dipole moment: Integration over all degrees of freedom }

Calculation of $N_{tot}$ dipole moments of Eq. (\ref{eq:dipole_long1}) implies integration over all degrees of freedom. In practice, the integration is accomplished in several steps. First, we integrate over radial coordinate $r$, calculating geometry dependent elements $m_o^\pm$ and $m_c$:
\begin{eqnarray}
\label{eq:mu_Jeff}
m_o^\pm\left({\cal Q}\right)=\pm\frac{1}{i\sqrt{2}}\int_0^\infty f_o^\pm(r) W_{ini}(r)\nu_{ini}^{-3/2}rdr, \\
m_c\left({\cal Q}\right)=\int_0^\infty W_c(r) W_{ini}(r;{\cal Q})\nu_{ini}^{-3/2}({\cal Q})rdr\nonumber.
\end{eqnarray}
In order to avoid strong energy-dependence of these elements, we do not include  factors ${\cal S}^{\dag phys}_{oi'}$ and $Z_{ci'}$ at this stage of the calculation. Therefore, the elements $m_o^\pm$ and $m_c$ are essentially independent of the photoionization energy. 

In Eq. (\ref{eq:mu_Jeff}) the Whittaker function $W_{ini}(r)$ is calculated for $l=0$ and the  effective quantum number $\nu_{ini}^{-3/2}({\cal Q})$ that depends of configuration and linked to the $3s$ potential  of H$_3$ as
\begin{equation}
\label{eq:nu_eff}
\nu_{ini}({\cal Q})=\frac{1}{\sqrt{2[V^{+}({\cal Q})-V^{3s}({\cal Q})]}},
\end{equation}
where $V^+({\cal Q})$ is the ionic potential and $V^{3s}({\cal Q})$ is the potential of the $3s$ state. Quantum defects needed to calculate functions $W_c(r)$ and $f_o^\pm(r)$ of final states are obtained from diagonalization of the scattering matrix $\cal S$. The integral of Eq. (\ref{eq:mu_Jeff}) is calculated numerically.

When evaluating Eq. (\ref{eq:mu_Jeff}), we must choose the principal quantum number $n$ for final states ($n$ is always 3 for the initial state). For a given energy $E$ of the photo-ionizing neutral molecule, $n_f$ is determined individually for every final state $f$  using ionization threshold energy $E_f^+$ of the channel $f$. If $E$ approaches closely to the threshold $E_f^+$, we take a large but finite $n$, typically $n=20$. Such approach is justified by the fact that the overlap in Eq. (\ref{eq:mu_Jeff}) depends weakly on  $n_f$ if $n_f$ is significantly larger than $n_i=3$. In principal, this procedure can be used for each photoionization energy $E$. Since the typical theoretical spectrum is calculated for typically more than $10^5$ energy points, this makes the evaluation of Eq. (\ref{eq:mu_Jeff}) at all energies $E$ expensive. To reduce the calculation time, we have adopted fixed values of $\nu_f$ and, therefore, fixed overlaps in Eq. (\ref{eq:mu_Jeff}) for several energies $E$. Since $m_j\left({\cal Q}\right)$ in Eq. (\ref{eq:mu_Jeff}) varies slowly with $E$, this approach provides much more rapid and sufficiently accurate values of the matrix elements in Eq. (\ref{eq:mu_Jeff}).

The next step in our evaluation of the dipole moments of Eq. (\ref{eq:dipole_long1}) is the integration over the vibrational coordinates $\cal Q$. The total vibrational function in the adiabatic approximation is represented as a product of hyper-radial and hyperangular components $\Psi^{vib}({\cal Q})=\psi_{v_1}(R)\Phi_{v_2l_2}(\theta, \phi)$. Knowing the wave functions $\Psi^{vib}({\cal Q}) $ and the $\cal Q$-dependent matrix moments of Eq.  (\ref{eq:mu_Jeff}), we can calculate the desired dipole matrix elements between initial and final vibrational states as follows:
\begin{eqnarray}
\label{eq:R_electronic}
R\left(\{v1,v2^{l_2}\}_j;\{v1,v2^{l_2}\}_{ini}\right)=\int d{\cal Q}\ \Psi^{vib*}_{\{v1,v2^{l_2}\}}({\cal Q})\ m_j\left( {\cal Q}\right)\ \Psi^{vib}_{ini}({\cal Q}).
\end{eqnarray}
where $m_j$ is $m_o^\pm$ or $m_c$.  In the same manner as for the scattering matrix, we evaluate this integral in two steps; first in the space of hyperangles, then in the space of the hyper-radius.

The next step is to evaluate the integration over angular coordinates in  (\ref{eq:dipole_long2}). To do this, we write explicitly all quantum numbers of the final and initial  wave functions, which is given by Eq. (\ref{eq:wf_long}), and we write the dipole moment $\tilde d_j$  as
\begin{eqnarray}
\label{eq:dipole_angular1}
\tilde d_{j}=\langle N^+_j,K^+_j;N_j,m_j \{v_1,v_2^{l_2}\}_j\big|\vec\epsilon\cdot \vec r\big|N^+_{ini},K^+_{ini};N_{ini},m_{ini} \{v_1,v_2^{l_2}\}_{ini}\rangle.
\end{eqnarray}
This expression can be represented in a form that is more suitable for our calculations. After some angular momentum algebra (for more details see the appendix) we derive the following formula for the dipole moment $\tilde d_j$.

\begin{eqnarray}
\label{eq:dipole_vector1}
\tilde d_{j}=\frac{1}{\sqrt{2N_j+1}}C_{1,0;1,m_{ini}}^{N_j,m_{ini}}\times\nonumber\\
\times\sum_{\Lambda_j} (-1)^{1-\Lambda_j}C_{l_j,-\Lambda_j;N_j,\Lambda_j}^{N^+_j,0} C_{1,\Lambda_j;1,0}^{N_j,\Lambda_j}  R\left(\{v_1,v_2^{l_2}\}_j;\ \{v_1,v_2^{l_2}\}_{ini}\right) .
\end{eqnarray}
It gives the dipoles moments $\tilde d_{j}$ in terms of vibrational matrix elements of Eq. \ref{eq:R_electronic}, which are calculated numerically.

In applying the present treatment to the interpretation of the photoionization experiments by Bordas {\it et al.} \cite{bordas91} and by Mistr\'\i k {\it et al.} \cite{mistrik00}, we consider only two different initial states specified by the set of quantum numbers $i=[\{v_1,v_2^{l_2}\}_{ini} l_{ini} \Lambda_{ini}]$. For the final state, $[\{v_1,v_2^{l_2}\}_j  l_j \Lambda_j]$, only the index $l_j=1$ is fixed --- we consider only $p$-wave final states.

\subsection{Final theoretical photoionization spectrum}

After the matrix elements $\tilde d_j$ have been obtained, the dipole transition moments $d_f$ are calculated using Eq. (\ref{eq:dipole_long1}). As we mentioned before, $\tilde d_j$ can be calculated once and used for all photoionization energies, provided the energy range of the entire calculated photoabsorption spectrum is not too extensive. However, the coefficients $\tilde v_{jf}$ must be recalculated at every final state energy of the theoretical photoionization spectrum.  Having calculated the dipole transition moments $d_f$, the total oscillator strength into the open ionization channels is then given by \cite{fano86}
\begin{equation}
\frac{df}{dE}=2\omega\sum_{f=1}^{N_o}\left| d_f\right|^2,
\end{equation}
where $\omega$ is the frequency of the laser light.

\subsection{Quantum defect parameters used in the calculation}

In order to construct the scattering matrix and dipole moment vector, we use the quantum defect parameters $\delta,\lambda,\mu_{\Lambda=0}({\cal Q})$, and $\mu_{\Lambda=\pm 1}$ determined by Mistr\'\i k {\it et al.}  \cite{mistrik00} from accurate {\it ab initio} calculations of potential-energy surfaces of H$_3$. As mentioned in our previous work \cite{kokoouline03a,kokoouline03b}, the parameters $\delta,\lambda$ determined in Ref. \cite{mistrik00} should be multiplied by factor $-\pi$ in order to correct the convention inconsistency in the definition of the reaction matrix $K$ in Ref. \cite{mistrik00}. The parameters $\delta,\lambda,\mu_{\Lambda=0}({\cal Q}),\mu_{\Lambda=\pm 1}$ are slightly different for different Rydberg states \cite{mistrik00}. In our calculation we use values obtained for $n=4$ Rydberg states: $\delta= -\pi\cdot 1090$ cm$^{-1},$ $\lambda=-\pi\cdot 12360$ cm$^{-1}$, $\mu_{\Lambda=\pm 1}=0.395$. The quantum defect $\mu_{\Lambda=0}({\cal Q})$ depends weakly on nuclear configuration $\cal Q$. We use $\mu_{\Lambda=0}({\cal Q})$ from Ref. \cite{mistrik00}. The ionization energies of 40 lowest states are taken from the same reference.

\section{Results and comparison with the experiments}
\label{sec:interpretation_of_spectr}

In the experiment by Bordas {\it et al.} \cite{bordas91}, the initial state of H$_3$ is the  state $i_1\equiv [\text{H}_3\ A_2'(tot.) \{00^0\}(N=1,K=0),\ 3s]$, the initial state in the experiment by Mistr\'\i k {\it et al.} \cite{mistrik00} differs only by the symmetric stretch vibrational quantum number $v_1$, which is the singly excited: $i_2\equiv [\text{H}_3\ A_2'(tot.) \{10^0\}(N=1,K=0),\ 3s(el.)]$. The energy difference between these two states is 3212.6 cm$^{-1}$. In the first experiment the energy region around the ground rovibrational level of the ion is probed by a tunable laser. The energy difference between the state $i_1$ and the ground rovibrational state $[\text{H}_3^+\ A_2'(ion) \{00^0\}(N^+=1,K^+=0)]$ of the ion is 12867.6 cm$^{-1}$.  In the second experiment \cite{mistrik00}, the energy region around the state with singly excited $v_1=1$ mode $[\text{H}_3^+\ A_2'(ion) \{10^0\}(N^+=1,K^+=0)]$ of the ion is probed. The energy difference between these two ionic levels, $[\text{H}_3^+\ A_2'(ion) \{00^0\}(N^+=1,K^+=0)]$  and $[\text{H}_3^+\ A_2'(ion) \{10^0\}(N^+=1,K^+=0)]$,  is 3176.06 cm$^{-1}$. In the present treatment the energy origin is set to the ground ionic state.

\begin{figure}[h]
\includegraphics[height=18cm]{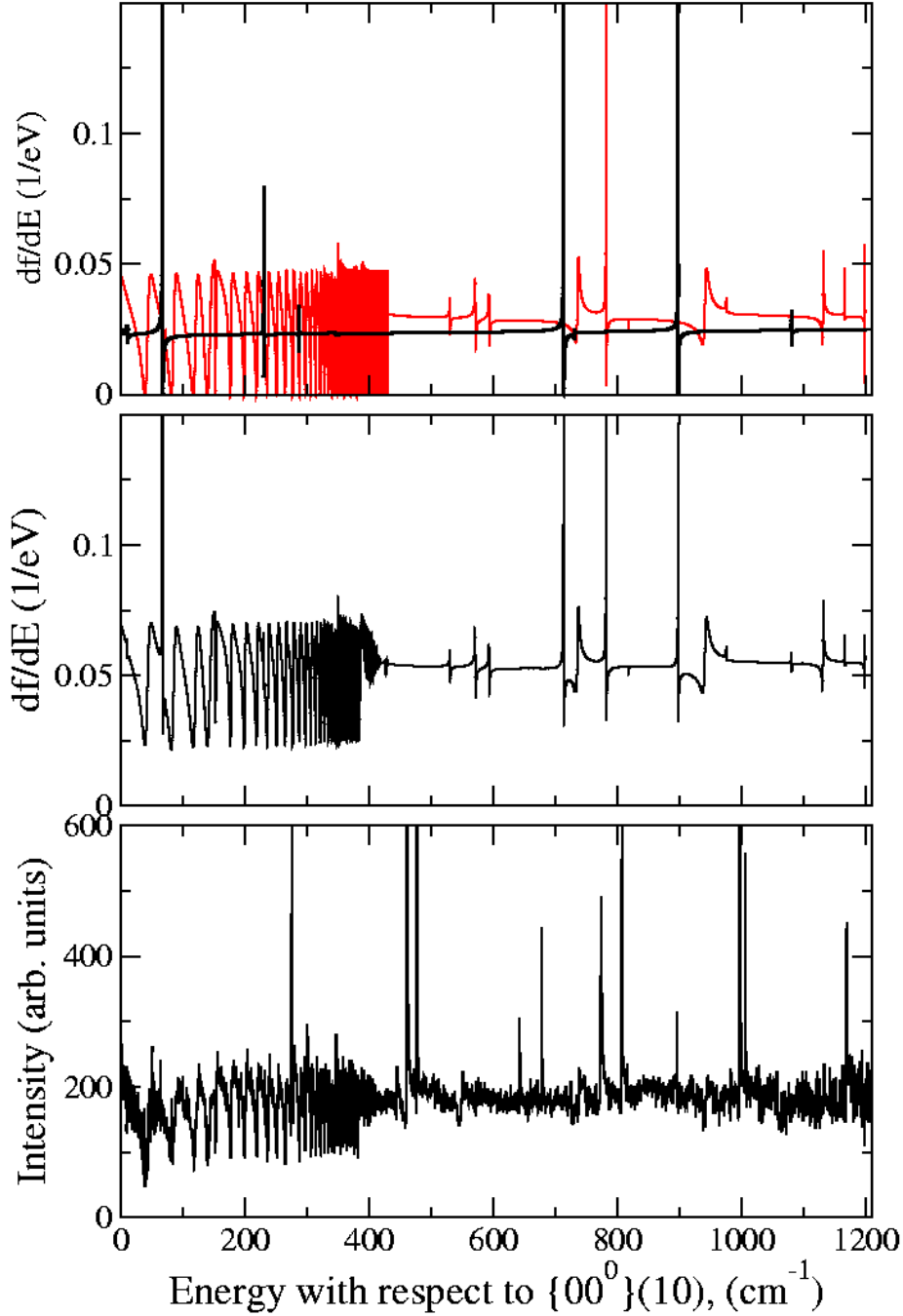}
\caption{\label{fig:comparison1_exp1}
Comparison of  calculated  photoionization spectra (upper and middle panel) and the experimental (lower panel) spectrum obtained by Bordas {\it et al.} \cite{bordas91}. In this experiment, the initial state of H$_3$ is $[3s^2A_2',\ \{00^0\}(1,0)]$. The upper panel shows theoretical spectra calculated separately for $N=0$ (black line) and  $N=2$ (gray line). The middle panel shows the sum result of the two spectra convolved with the experimental resolution width 0.15 cm$^{-1}$. Therefore, the spectrum in the middle panel should be compared with the experimental spectrum. Here and in all following figures, the energy is related to the ground rovibrational level of H$_3^+(A_2')$, $\{00^0\}(1,0)$.
}
\end{figure}

\begin{figure}[h]
\includegraphics[height=18cm]{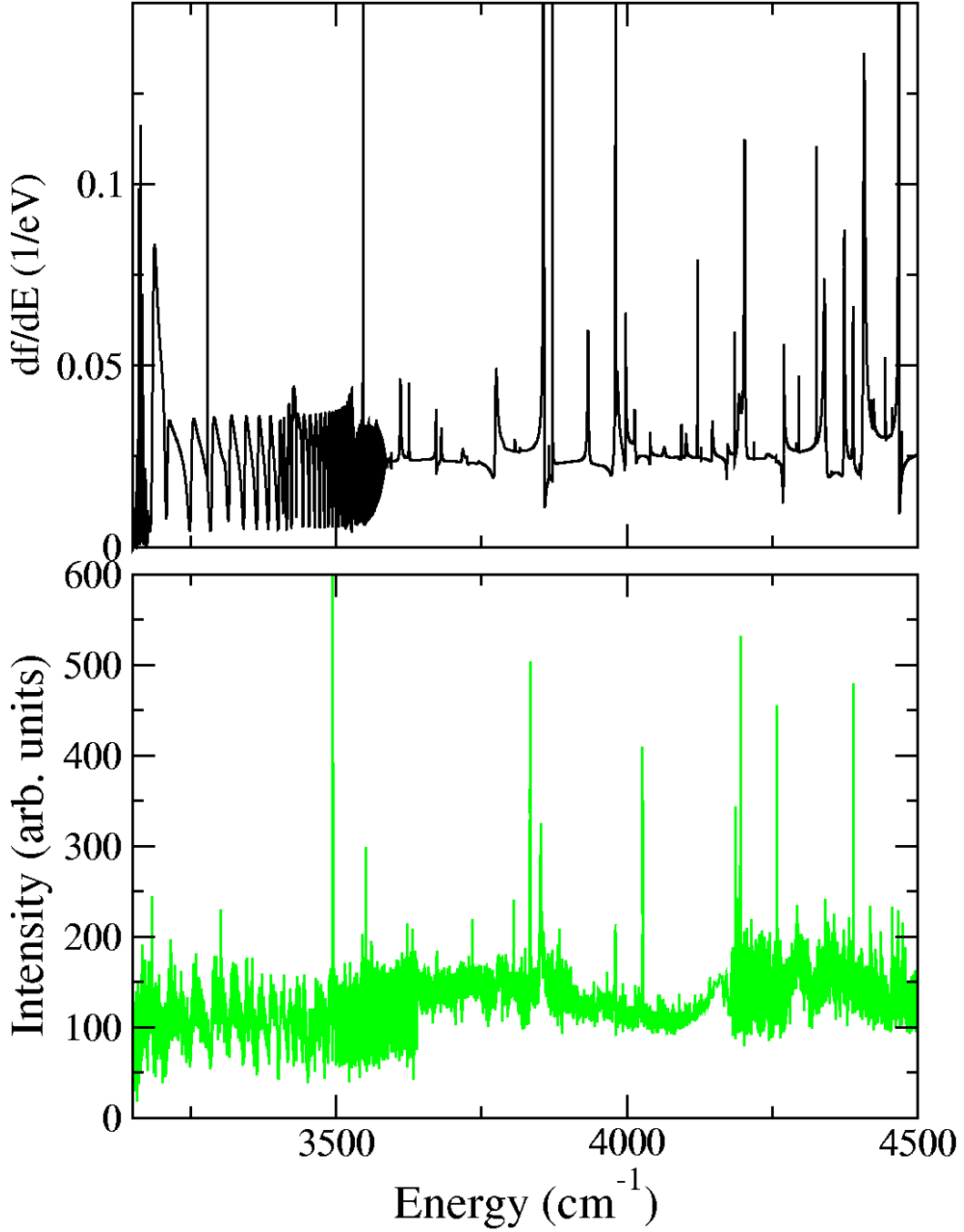}
\caption{\label{fig:comparison1_exp2}
Comparison of the theoretical  (upper panel) spectrum with the observed one (lower panel) from experiment of Ref. \cite{mistrik00}. In this experiment, the initial state differs from the experiment by Bordas {\it et al.}: It is $[3s^2A_2',\ \{10^0\}(1,0)]$. In constructing the theoretical spectrum, we have combined the spectra for $N=0$ and $N=2$ according to experimental conditions (see text). The final spectrum is convolved with the experimental resolution width of 0.2 cm$^{-1}$.
}
\end{figure}

Figures (\ref{fig:comparison1_exp1}) and (\ref{fig:comparison1_exp2}) shows experimental and calculated spectra for the two experiments. The overall agreement is quite good. Below we give a more detailed discussion of the comparison between the experiments and our calculation.

In constructing the theoretical spectrum, we have combined the spectra for $N=0$ and $N=2$ according to the experimental conditions. Specifically, to compare our theoretical results with the experiment by Bordas {\it et al.}, we have summed up the separate theoretical spectra  for $N=0$  and  $N=2$. To compare with the experiment by Mistr\'\i k {\it et al.} we have accounted for a fixed angle of $60^o$ between the linear polarization vectors of two the lasers  used in the  experiment. We used the prescription of Ref. \cite{mistrik00}, according to which the final spectrum $\frac{df}{dE}$ is constructed as
\begin{equation}
\frac{df}{dE}=\frac{1}{4}\frac{df_0}{dE}+\frac{13}{16}\frac{df_2}{dE},
\end{equation}
where $\frac{df_0}{dE}$ and $\frac{df_2}{dE}$ are the spectra calculated for $N=0$  and  $N=2$, respectively.

The energy regions accessible in the experiments have three qualitatively different regimes, namely the discrete, Beutler-Fano, and continuum regimes. 

\begin{figure}[h]
\includegraphics[width=17cm]{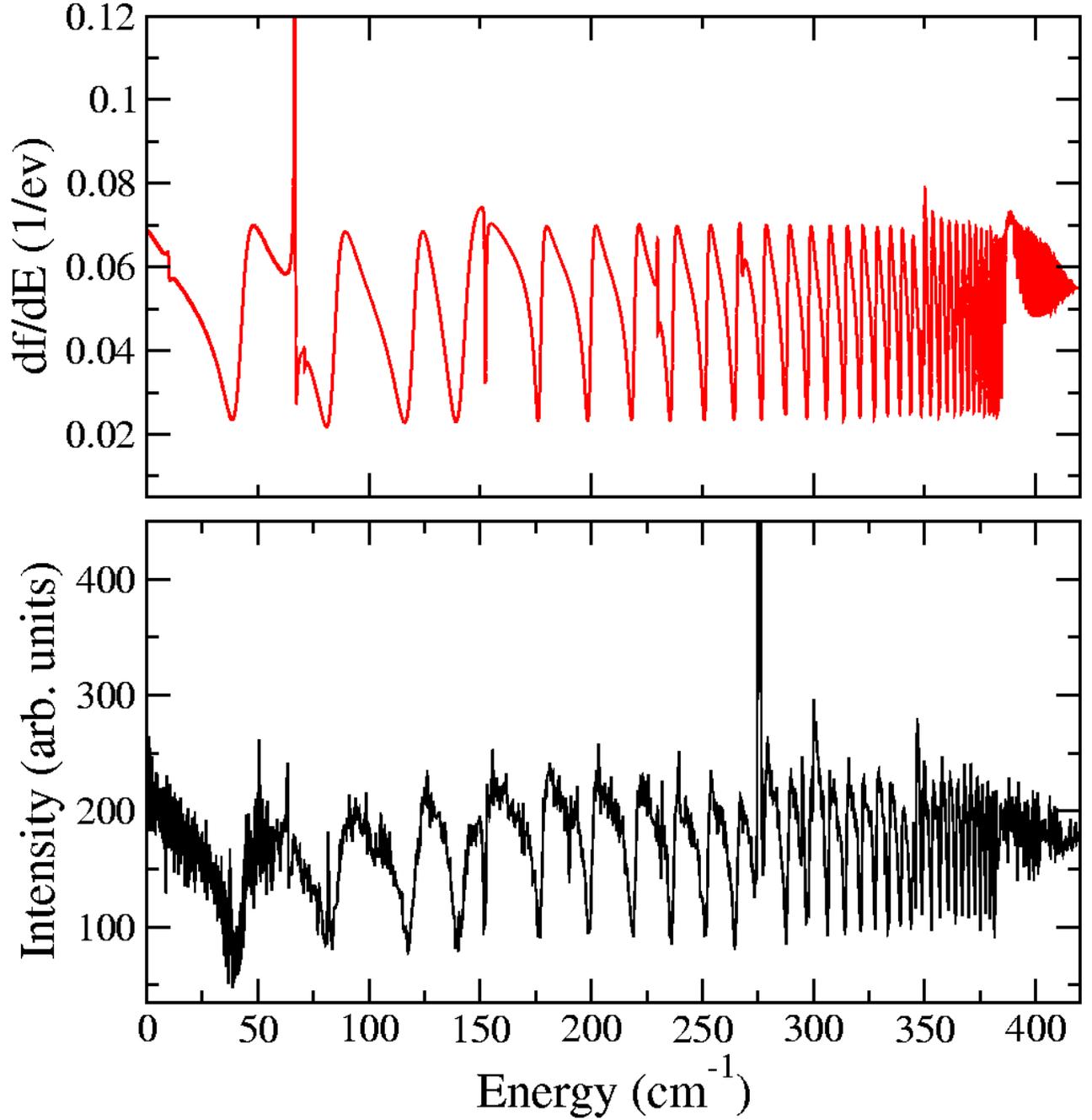}
\caption{\label{fig:comparison2_exp1}
Comparison between theoretical (upper panel) and experimental (lower panel) results: the Beutler-Fano part of the spectrum shown in Fig. \ref{fig:comparison1_exp1}. This energy region corresponds to energies between $\{00^0\}N^+=3$ and $\{00^0\}N^+=1$ ionic rotational levels. The series of wide resonances are due to the fast rotational autoionization of states with $N^+=3$  to an open continuum of the $N^+=1$ rotational level. Interlopers at 70 cm$^{-1}$ and 150 cm$^{-1}$ are two  examples of resonances playing an important role in DR of H$_3^+$.
}
\end{figure}

\begin{figure}[h]
\includegraphics[width=17cm]{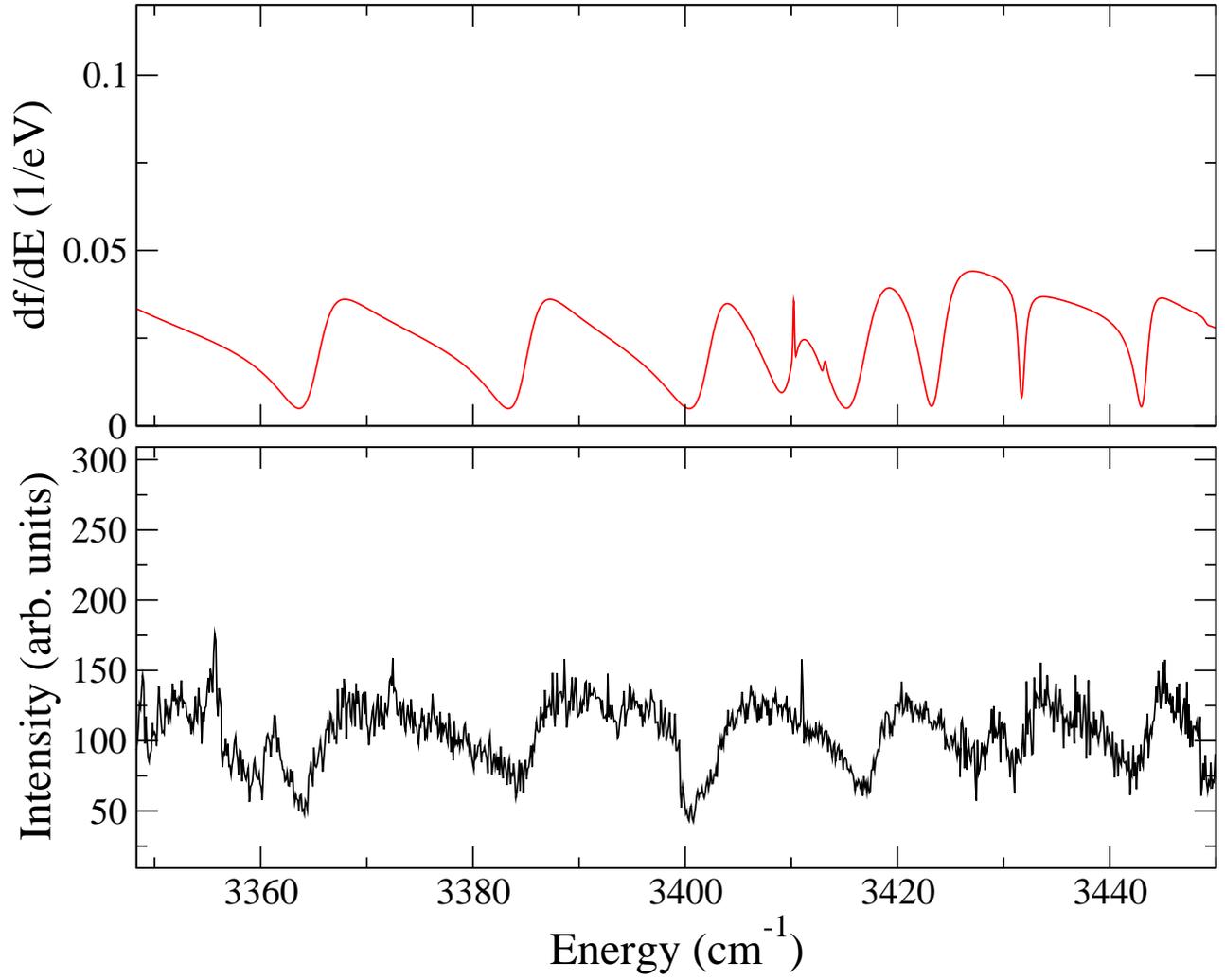}
\caption{\label{fig:comparison3_exp2}
Theoretical (upper panel) and experimental (lower panel, from Ref. \cite{mistrik00})  Beutler-Fano spectra. Theory reproduces the most of experimental features.
}
\end{figure}

The Beutler-Fano regime arises at energies between two different rotational levels associated with the same vibrational state. For both experiments, this region occurs at final state energies between the $[\{v_10^0\}(N^+=1,K^+=0)]$ and $[\{v_10^0\}(N^+=3,K^+=0)]$ states of the ion, where $v_1=0$ for the experiment of Ref. \cite{bordas91} and $v_1=1$ for the experiment of Ref.  \cite{mistrik00}. Autoionization in this region usually occurs quite rapidly compared to autoionization in other energy ranges. Consider, for example, an electron is excited into a Rydberg state attached to the $[\{v_10^0\}(N^+=3,K^+=0)]$ ionic level, at a total energy above  $\{v_10^0\}(N^+=1,K^+=0)]$. Since both rotational levels have the same vibrational excitation, the corresponding Franck-Condon overlap between the Rydberg state and a continuum state of the  $\{v_10^0\}(N^+=1,K^+=0)]$ level is very favorable, which generates a large autoionization width, as is evident in the Beutler-Fano regions of both experiments. Figure \ref{fig:comparison2_exp1} shows a detailed  comparison between theory and the experiment \cite{bordas91} for the Beutler-Fano region, and Fig. \ref{fig:comparison3_exp2} presents a detailed comparison with the second experiment \cite{mistrik00}. The agreement between theory and experiment is  good, which is evidence supporting the approximations we have adopted in our theoretical description.

\begin{figure}[h]
\includegraphics[width=17cm]{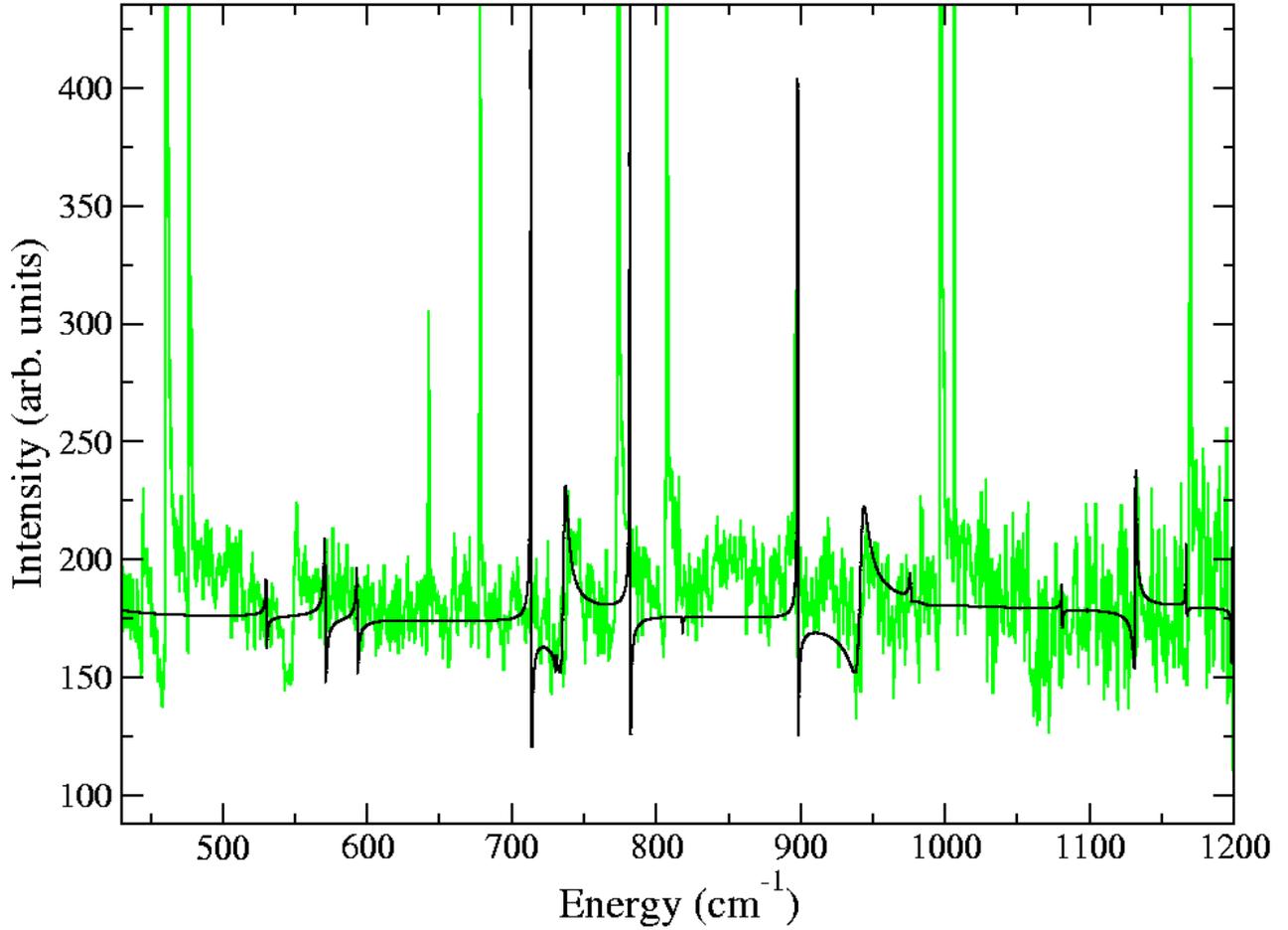}
\caption{\label{fig:comparison4_exp1}
A part of the theoretical (dark line) and the experimental (gray line) spectra (see Fig. \ref{fig:comparison1_exp1}) corresponding to the H$_3$ continuum above the $\{00^0\}(30)$ rovibrational level of H$_3^+$.  Relatively wide resonances at 740  cm$^{-1}$ and 950 cm$^{-1}$ are produced by the Jahn-Teller coupling between electronic and vibrational motion of H$_3$. Although the overall agreement between theory and experiment is better than in previous theoretical studies \cite{bordas91,stephens94,stephens95},  several experimental features still exist that are not reproduced in the present approach.
}
\end{figure}

\begin{figure}[h]
\includegraphics[width=17cm]{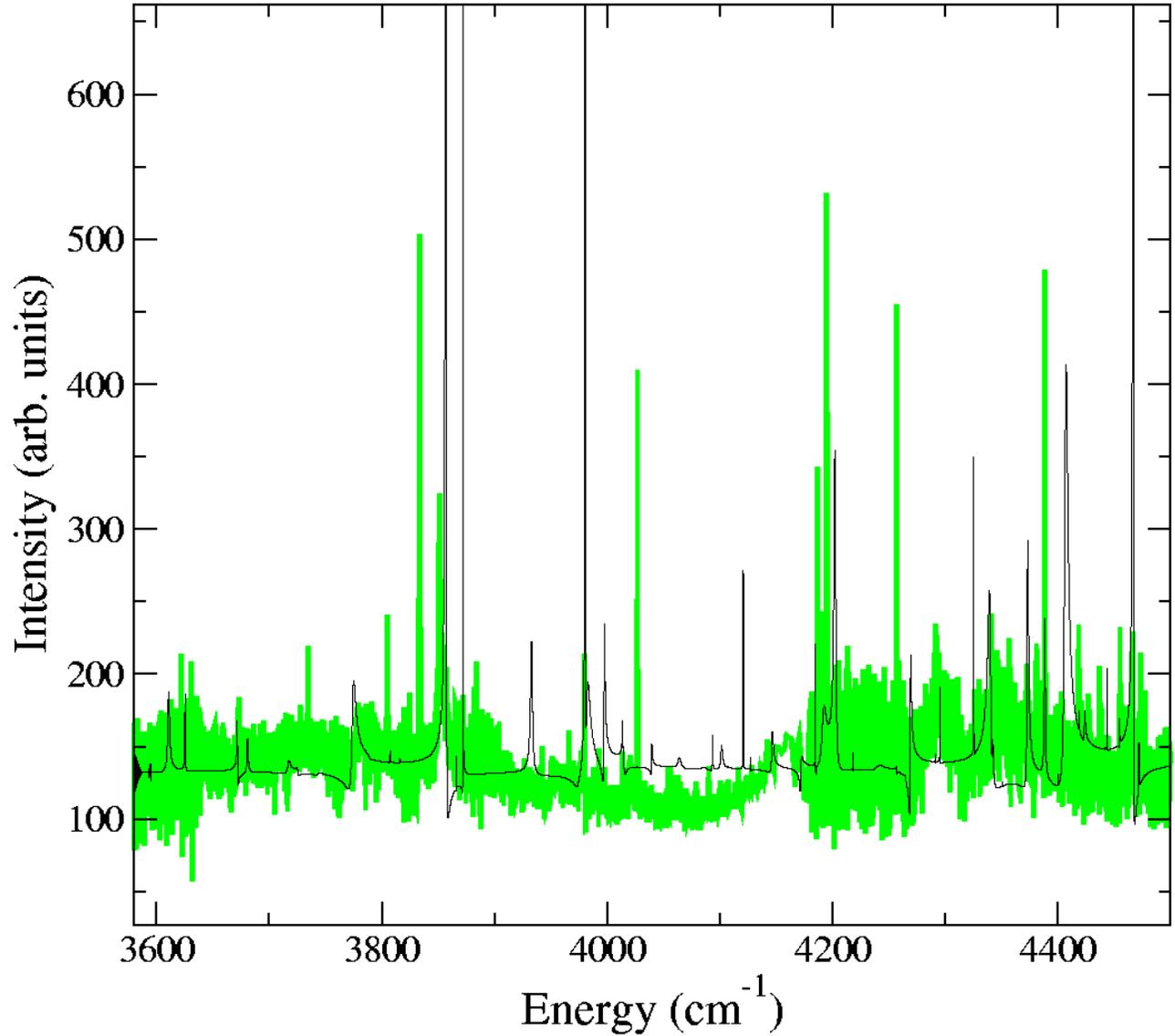}
\caption{\label{fig:comparison2_exp2}
Theoretical (dark line) and experimental (gray line) spectra (see Fig. \ref{fig:comparison1_exp2}) corresponding to the H$_3$ continuum above the $\{10^0\}(30)$ rovibrational level of H$_3^+$. Similar to Fig. \ref{fig:comparison4_exp1}, the agreement between theory and experiment is better than in the previous theoretical study \cite{mistrik00}, but there are several uninterpreted experimental resonances.
}
\end{figure}

The continuum regime is situated above the corresponding Beutler-Fano energy range. Generally, autoionization is much slower in such regions. Figure \ref{fig:comparison4_exp1} compares the calculated spectrum with the experiment of Ref. \cite{bordas91}. We draw attention to two broad resonances around 740 cm$^{-1}$ and 950 cm$^{-1}$. Not only do these nicely reproduce the experimental spectrum, but, importantly, they are caused by Jahn-Teller coupling. However, in the previous experimental and theoretical studies, these features were ignored, probably because they were construed as noise. The present calculation suggests that these are real resonances, broadened by a strong interaction between rotational and vibrational degrees of freedom.  Figure \ref{fig:comparison2_exp2} compares the present calculation with the results of Ref. \cite{mistrik00}. In this figure we would like to note two other features. The experimental data is rather noisy, but an inspection of the calculated and experimental spectra suggests that the calculated resonances appearing around 3770 cm$^{-1}$ and around 3850 cm$^{-1}$ correspond to broad observed resonances. Again, the large widths of the two resonances are caused by Jahn-Teller coupling.

\begin{figure}[h]
\includegraphics[width=17cm]{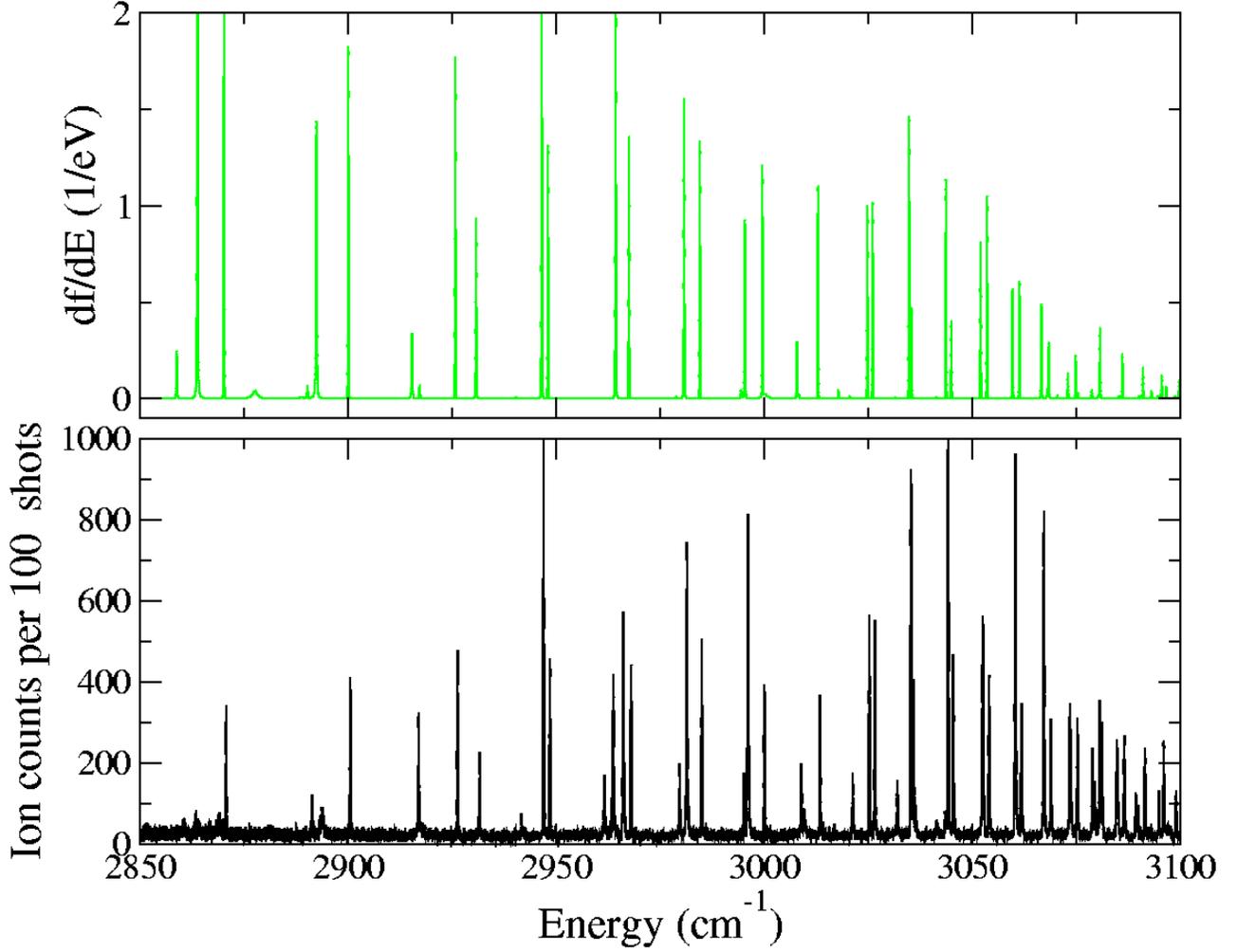}
\caption{\label{fig:comparison4_exp2}
Theoretical (upper panel) and experimental (lower panel) \cite{mistrik00} spectra of H$_3$ in the quasi-discrete energy region below the $\{10^0\}(10)$ rovibrational level of H$_3^+$. The initial state for the dipole transition (H$_3$ photoexcitation) is $[3s^2A_2',\ \{10^0\}(1,0)]$. This region is referred as quasi-discrete since states, populated starting from $[3s^2A_2',\ \{10^0\}(1,0)]$, are only weakly interacting with lower ionic states with $v=\{00^0\}$. 
}
\end{figure}

The discrete regime is the energy range where autoionization of Rydberg states is energetically forbidden. However, we will follow the convention proposed in Ref. \cite{mistrik00} and will call the region below $[\text{H}_3^+\ A_2'(ion) \{00^0\}(N^+=1,K^+=0)]$ as discrete too when we refer to  $i_2$ as the initial state.  This convention is justified by the fact that autoionization of such states is slow, owing to an unfavorable Franck-Condon overlap of these states with the ground ionic state. Figure \ref{fig:comparison4_exp2} shows a comparison of experimental \cite{mistrik00} and calculated spectra for the discrete spectrum of the $i_2$ initial state.

Although the overall agreement between theory and experiment is still good overall, there are some resonances in the experimental spectrum that absent from the calculated spectrum. Some of these missing resonances could conceivably be caused by an influence of the $d$ electronic wave. This possibility was demonstrated by  Mistr\'\i k \cite{mistrik01thesis}. For example, an additional resonance in the experimental spectrum around 2970 cm$^{-1}$ appears to be caused by an $nd\sigma$ Rydberg electron. 

\section{Conclusion}
\label{sec:concl}

In the present study, we propose an updated theoretical method for treatment of photoionization in the H$_3$ molecule. The main engine of the method, MQDT including the Jahn-Teller coupling and the rovibrational frame transformation, is the same as in studies by Stephens and Greene \cite{stephens94,stephens95}. However, we have improved beyond Refs. \cite{mistrik00,stephens94,stephens95} in the treatment of the symmetry issues associated with different degrees of freedom. We have proposed a new and more efficient symmetrization procedure that accounts for the symmetries of nuclear spin as well as the rotational and vibrational parts of the total wave function.  Another improvement is that we include into the description of  H$_3$ photoionization the possibility that the system may break apart by dissociation in addition to ionization. To our knowledge, this is the first time that photodissociation has been included in its competition with photoionization for richly resonant Rydberg spectrum of a triatomic molecule. Although in this system, predissociation apparently does not play an important role for the overall spectra under consideration, it must be important, at least, for some Rydberg states of H$_3$. In fact, the importance of the dissociation channel was recently demonstrated experimentally \cite{mistrik01}. In a future study, hopefully, we will investigate theoretically the predissociation of such Rydberg states. The inclusion of predissociation  could also be important in studies of other triatomic molecules. Note that as in Ref. \cite{kokoouline03a,kokoouline03b}, we have corrected the earlier inconsistency in the reaction matrix convention employed in Refs. \cite{mistrik00,stephens94,stephens95,kokoouline01} which results in a multiplication of the Jahn-Teller parameters $\lambda$ and $\delta$ employed in those references by the factor of $-\pi$. In the present study, we have also improved the dipole moment calculation, and reformulated it in a form that should be particularly accurate for the photoionization of a Rydberg molecular initial state. In contrast to most previous studies of H$_3$ photoionization,\cite{mistrik00,stephens94,stephens95} the present treatment makes use of the adiabatic hyperspherical approach for the vibrational motion of the nuclei in the H$_3^+$ ion and in the H$_3$ initial state being photoionized. Owing to the adiabatic approximation, this approach should be {\it a priori} less accurate than the one used in Refs. \cite{mistrik00,stephens94,stephens95}, but the loss of accuracy due to this approximation seems to be small. The possibility of constructing a unified theoretical description of H$_3$ photoionization and H$_3^+$ dissociative recombination is an attractive feature of this approach.

We have obtained good agreement with both photoionization experiments \cite{bordas91,mistrik00} and with many of the spectra obtained in previous theoretical studies \cite{mistrik00,stephens94,stephens95}. In some regions the agreement with the experiments is even better in the present study than it was in Refs. \cite{mistrik00,stephens94,stephens95}. We attribute this to the correction of the aforementioned error in determination of the Jahn-Teller coupling parameters $\lambda$ and $\delta$. Although we have been able to reproduce most of the observed resonance features, both in position and in shape, there remain several features in the experimental spectra that are not described by our treatment. One possible explanation of the discrepancy between theory and experiment is the influence of $nd$ Rydberg states, as proposed by  Mistr\'\i k \cite{mistrik01thesis}, because we have not taken these states into consideration.

In conclusion, we would reiterate that we have developed a new theoretical method that can be applied to a unified treatment of photoionization and dissociative recombination for molecules of the $D_{3h}$ symmetry group. Our application  to the H$_3$ molecule shows good general agreement with existing experimental observations of H$_3$ photoionization. The method has also been shown in our previous study \cite{kokoouline03a,kokoouline03b} to give good agreement with measurements of H$_3^+$ dissociative recombination.

\section{Appendix}
In the appendix we provide a detailed derivation of Eq. \ref{eq:dipole_vector1} starting from Eq. \ref{eq:dipole_angular1}.

Suppose the laser light is linearly polarized in the $Z$ direction in LS. The covariant spherical tensor components of the polarization vector are then $(\epsilon_{+1},\epsilon_{0},\epsilon_{-1})=(0,1,0)$. The vector $\vec r$ in spherical coordinates is \cite{varshalovich}
\begin{eqnarray}
\label{eq:dipole1}
r_\mu=\sqrt{\frac{4\pi}{3}}|r|Y_{1,\mu}(\theta ,\phi) \text{ with } \mu=0,\pm 1 .
\end{eqnarray}
Then, the scalar product $\vec\epsilon\cdot \vec r$ in LS is
\begin{eqnarray}
(\vec\epsilon\cdot \vec r)_{LS}=\sqrt{\frac{4\pi}{3}}|r| Y_{1,0}(\theta ,\phi).
\end{eqnarray}
The expression of this product in MS is obtained by an appropriate coordinate rotation,
\begin{eqnarray}
(\vec\epsilon\cdot \vec r)_{MS}=\sqrt{\frac{4\pi}{3}}|r| \sum_{\Lambda'} Y_{1,\Lambda '}(\theta ',\phi ')\left[{\cal D}_{0,\Lambda'}^1(\alpha,\beta,\gamma)\right]^*.
\end{eqnarray}
Accounting Eqs. (\ref{eq:wf_short2}) and (\ref{eq:trans2}), the expression for the amplitude of the dipole transition of Eq. (\ref{eq:dipole_angular1}) becomes
\begin{eqnarray}
\tilde d_{j}=\sqrt{\frac{4\pi}{3}}\sum_{\Lambda'} \langle N^+_j,K^+_j;N_j,m_j\{v_1,v_2^{l_2}\}_j\big|Y_{1,\Lambda'}(\theta ',\phi ')\left[{\cal D}_{0,\Lambda'}^1(\alpha,\beta,\gamma)\right]^*\times\nonumber\\
\times\big|N^+_{ini},K^+_{ini};N_{ini},m_{ini} \{v_1,v_2^{l_2}\}_{ini}\rangle=\nonumber\\
=\sqrt{\frac{4\pi}{3}}\sum_{\Lambda',\Lambda_j}C_{l_j,-\Lambda_j;N_j,K_j}^{N^+_j,K^+_j}(-1)^{l_j-\Lambda_j}\big<{\cal R}(N_j,m_j,K_j;\alpha  \beta  \gamma)Y_{l_j\Lambda_j}(\theta',\varphi')  Y_{1,\Lambda'}(\theta ',\phi ')\times\nonumber\\
\times\left[{\cal D}_{0,\Lambda'}^1(\alpha\beta\gamma)\right]^*{\cal R}(N_{ini},m_{ini},K_{ini};\alpha  \beta  \gamma)Y_{l_{ini}\Lambda_{ini}}(\theta',\varphi')\big>_{\alpha\beta\gamma\theta'\varphi'}C_{l_{ini},-\Lambda_{ini};N_{ini},K_{ini}}^{N^+_{ini},K^+_{ini}}\times\nonumber\\
\times (-1)^{l_{ini}-\Lambda_{ini}} R\left(\{v_1,v_2^{l_2}\}_j;\ \{v_1,v_2^{l_2}\}_{ini}\right).
\end{eqnarray}
If the initial electronic state is $3s$, then $Y_{l_{ini}\Lambda_{ini}}=Y_{0,0}=1/\sqrt{4\pi}$. The integral over electronic angles $\theta',\varphi'$ is then trivial, and the relevant angular matrix element is
\begin{eqnarray}
\big<Y_{l_j\Lambda_j}(\theta',\varphi')|Y_{1,\Lambda'}(\theta ',\phi ')|Y_{l_{ini}\Lambda_{ini}}(\theta',\varphi')\big>_{\theta'\varphi'}=\frac{\delta_{\Lambda_j,\Lambda'}}{2\sqrt{\pi}}.
\end{eqnarray}
The integral over $\alpha,\beta$, and $\gamma$ angles is evaluated using the addition theorem for Wigner functions and their normalization properties \cite{varshalovich}:
\begin{eqnarray}
\big<{\cal R}(N_j,m_j,K_j;\alpha  \beta  \gamma)\big|\left[{\cal D}_{0,\Lambda'}^1(\alpha\beta\gamma)\right]^*\big|{\cal R}(N_{ini},m_{ini},K_{ini};\alpha  \beta  \gamma)\big>_{\alpha\beta\gamma}=\nonumber\\
=\sqrt{\frac{(2N_j+1)(2N_j+1)}{(8\pi^2)^2}}\left[\big< D^{N_j}_{m_j,K_j}\big|D_{0,\Lambda'}^1\big|D^{N_{ini}}_{m_{ini},K_{ini}}\big>_{\alpha\beta\gamma}\right]^*=\nonumber\\
=\sqrt{\frac{(2N_j+1)(2N_{ini}+1)}{(8\pi^2)^2}}\sum_{N'}\left[\big< D^{N_j}_{m_j,K_j}\big| D^{N'}_{m_{ini},K_{ini}+\Lambda'}C_{1,0;N_{ini},m_{ini}}^{N',m_{ini}}C_{1,\Lambda';N_{ini},K_{ini}}^{N',K_{ini}+\Lambda'}    \big>_{\alpha\beta\gamma}\right]^*=\nonumber\\
=\sqrt{\frac{2N_{ini}+1}{2N_j+1}}C_{1,0;N_{ini},m_{ini}}^{N_j,m_{ini}}C_{1,\Lambda';N_{ini},K_{ini}}^{N_j,K_{ini}+\Lambda'}\delta_{N_j,N'}\delta_{m_j,m_{ini}}\delta_{K_j,K_{ini}+\Lambda'}=\\
=\sqrt{\frac{2N_{ini}+1}{2N_j+1}}C_{1,0;N_{ini},m_{ini}}^{N_j,m_{ini}}C_{1,\Lambda';N_{ini},K_{ini}}^{N_j,K_{ini}+\Lambda'}\nonumber.
\end{eqnarray}
The matrix element then becomes 
\begin{eqnarray}
\tilde d_{j}=\frac{1}{\sqrt{3}}\sqrt{\frac{2N_{ini}+1}{2N_j+1}}\times\nonumber\\
\sum_{\Lambda_j}C_{l_j,-\Lambda_j;N_j,K_j}^{N^+_j,K^+_j} C_{1,0;N_{ini},m_{ini}}^{N_j,m_{ini}}C_{1,\Lambda_j;N_{ini},K_{ini}}^{N_j,K_{ini}+\Lambda_j}\delta_{m_j,m_{ini}}\delta_{K_j,K_{ini}+\Lambda_j}   C_{l_{ini},-\Lambda_{ini};N_{ini},K_{ini}}^{N^+_{ini},K^+_{ini}}\times\nonumber\\
(-1)^{1-\Lambda_j}  R\left(\{v_1,v_2^{l_2}\}_j;\ \{v_1,v_2^{l_2}\}_{ini}\right).
\end{eqnarray}
Using the fact that $N_{ini}=N_{ini}^+=1,\ K_{ini}=K_{ini}^+,\ m_{ini}=m_{ini}^+=m_j,\ K_j=K_{ini}+\Lambda_j$ and also the requirement $K_{ini}^+=0$, which arises from  symmetry restrictions on the initial rovibrational state, the last relation simplifies in this case to:
\begin{eqnarray}
\label{eq:dipole_vector2}
\tilde d_{j}=\frac{1}{\sqrt{2N_j+1}}C_{1,0;1,m_{ini}}^{N_j,m_{ini}}\times\nonumber\\
\times\sum_{\Lambda_j} (-1)^{1-\Lambda_j}C_{l_j,-\Lambda_j;N_j,\Lambda_j}^{N^+_j,0} C_{1,\Lambda_j;1,0}^{N_j,\Lambda_j}  R\left(\{v_1,v_2^{l_2}\}_j;\ \{v_1,v_2^{l_2}\}_{ini}\right).
\end{eqnarray}
The last formula is Eq. \ref{eq:dipole_vector1}.

\begin{acknowledgments}
This work has been supported in part by NSF, by the DOE Office of Science, and by an allocation of NERSC supercomputing resources. The authors are grateful to  H.~Helm, M.~Larsson, J.~Stephens, and I.~Mistr\'\i k for fruitful discussions and to E. Hamilton and B. Esry for assistance.
\end{acknowledgments}

\begin{figure}[h]
\includegraphics[width=15cm]{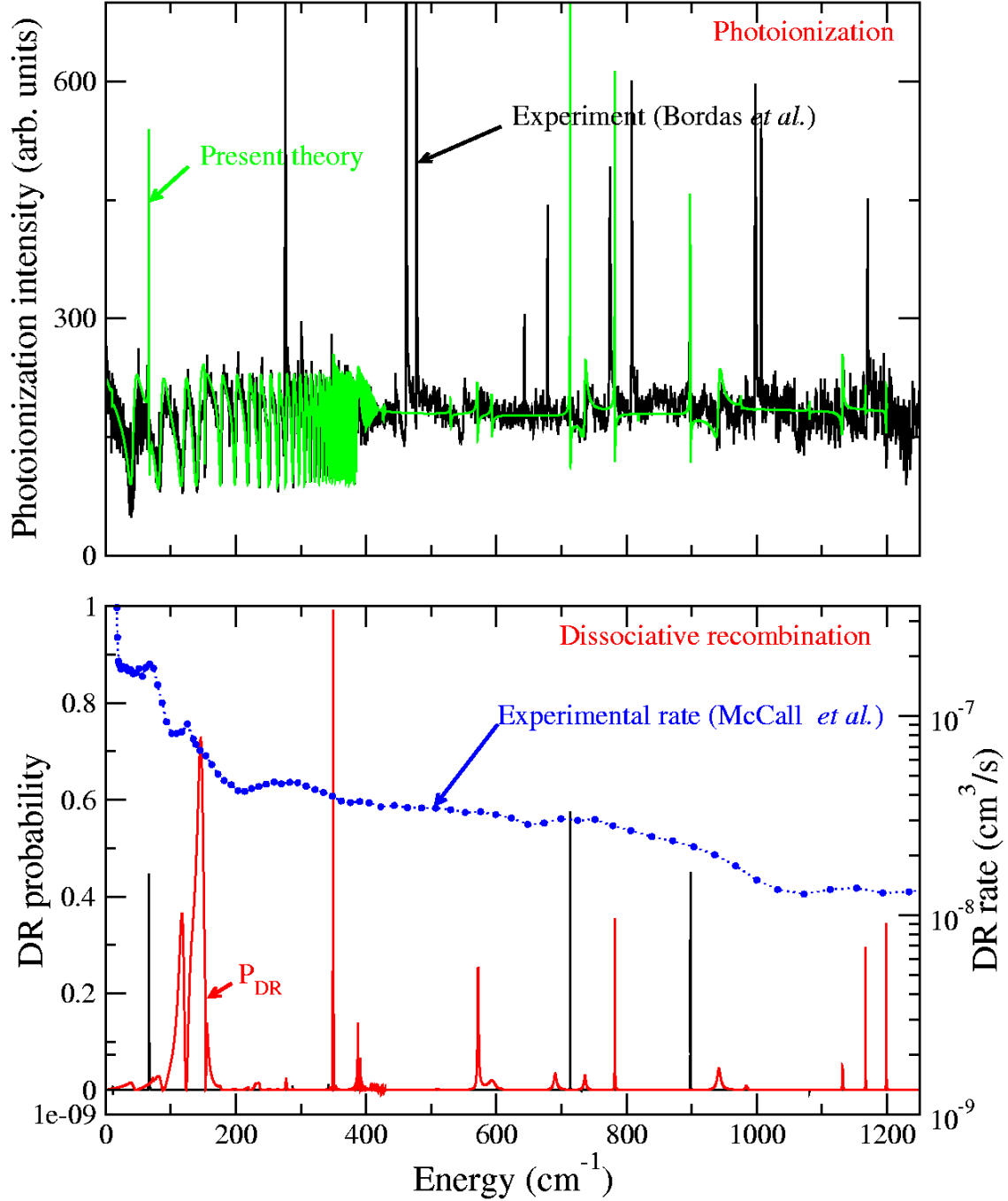}
\caption{\label{DR_PI_comp}
The figure demonstrates  a correlation between photoionization (upper panel) and dissociative recombination (lower panel) spectra obtained theoretically and in the experiment. The upper panel shows both experimental \cite{bordas91} and theoretical spectra, similar to Fig. \ref{fig:comparison1_exp1}. The lower panel gives an experimental DR rate obtained in Ref. \cite{mccall03}, which is shown as circles. Grey line in lower panel gives theoretical probability of DR process, calculated for the total $A_2'$ symmetry of H$_3^+$. This symmetry corresponds to the total symmetry of H$_3$ in the both photoionization experiments considered in this study.
}
\end{figure}

\end{document}